\documentclass[amsmath,amssymb,twocolumn,superscriptaddress,floatfix,preprintnumbers, prb]{revtex4}
\usepackage[pdftex]{graphicx}
\usepackage{subfigure}
\usepackage{dcolumn}
\usepackage{bm}
\usepackage[usenames,dvipsnames,svgnames]{xcolor}
\usepackage[colorlinks]{hyperref}
\hypersetup{
    linkcolor=BrickRed,
    urlcolor=blue,
    citecolor=MidnightBlue,
}

\RequirePackage[normalem]{ulem} %DIF PREAMBLE
\RequirePackage{color}\definecolor{RED}{rgb}{1,0,0}\definecolor{BLUE}{rgb}{0,0,1} %DIF PREAMBLE
\newcommand{\abs}[1]{\vert #1 \vert}
\newcommand{\bra}[1]{\langle #1 \vert}
\newcommand{\ket}[1]{\vert #1 \rangle}
\newcommand{\braket}[1]{\langle #1 \rangle}
\newcommand{\expect}[1]{\langle #1 \rangle}

\renewcommand{\vec}[1]{{\mathbf #1}}

\newcommand{\dn}[0]{\downarrow}
\newcommand{\up}[0]{\uparrow}

\newcommand{\plane}[2]{$#1$\nobreakdash-$#2$~plane}
\newcommand{\xyplane}{\plane{x}{y}}

\newcommand{\UIUCECE}[0]{\affiliation{Department of Electrical and Computer Engineering, University of Illinois at Urbana-Champaign, Urbana, IL 61801, USA}}
\newcommand{\UIUCPHYS}[0]{\affiliation{Department of Physics, University of Illinois at Urbana-Champaign, Urbana, IL 61801, USA}}
\newcommand{\UIUCMNTL}[0]{\affiliation{Micro and Nanotechnology Laboratory, University of Illinois, 208 N. Wright Street, Urbana IL 61801, USA}}

\def\be{\begin{equation}}
\def\ee{\end{equation}}
\def\bea{\begin{eqnarray}}
\def\eea{\end{eqnarray}}

\bibpunct{(}{)}{;}{s}{,}{,}

\begin{document}

\title{Topological superconductivity in an ultrathin, magnetically-doped topological insulator proximity coupled to a conventional superconductor}

%\author{nanotransport group}\UIUCECE\UIUCPHYS\UIUCMNTL
 \author{Youngseok Kim}\UIUCECE\UIUCMNTL
 \author{Timothy M. Philip}\UIUCECE\UIUCMNTL
 \author{Moon Jip Park}\UIUCMNTL\UIUCPHYS
 \author{Matthew J. Gilbert}\UIUCECE\UIUCMNTL
\date{\today}

\begin{abstract}
As a promising candidate system to realize topological superconductivity, the system of a 3D topological insulator (TI) grown on top of the $s$-wave superconductor has been extensively studied. 
To access the topological superconductivity experimentally, the 3D TI sample must be thin enough to allow for Cooper pair tunneling to the exposed surface of TI. 
The use of magnetically ordered dopants to break time-reversal symmetry may allow the surface of a TI to host Majorana fermion, which are believed to be a signature of topological superconductivity. 
In this work, we study a magnetically-doped thin film TI-superconductor hybrid systems. Considering the proximity induced order parameter in thin film of TI, 
we analyze the gap closing points of the Hamiltonian and draw the phase diagram as a function of relevant parameters: the hybridization gap, Zeeman energy, and chemical potential of the TI system. Our findings provide a useful guide in choosing relevant parameters to facilitate the observation of topological superconductivity in thin film TI-superconductor hybrid systems. In addition, we further perform numerical analysis on a TI proximity coupled to a $s$-wave superconductor and find that, due to the spin-momentum locked nature of the surface states in TI, the induced $s$-wave order parameter of the surface states persists even at large magnitude of the Zeeman energy. 
\end{abstract}
\maketitle

\section{Introduction} 
A particularly interesting direction within the landscape of topological systems is to realize topological superconductivity. Within condensed matter physics there has been a significant effort to find systems that exhibit topological superconductivity as such systems are predicted to harbor the, heretofore, elusive Majorana fermions\cite{Fu2008,Lutchyn2010,Mourik2012,Nadj-perge2014}. There have been a number of proposals that have been predicted to realize topological superconductivity, and these proposals may be grouped into two: (i) unconventional superconducting materials such as Sr$_2$RuO$_4$\cite{Jang2011,Kallin2009,Kallin2012} or doped superconducting materials such as Cu$_{x}$Bi$_{2}$Se$_{3}$\cite{Hor2010a,Fu2010,Tafti2013} and (ii) proximity-coupled system comprised of a conventional superconductor and a system such as strongly spin-orbit coupled semiconductors\cite{Sau2010}, magnetic adatoms\cite{Nadj-Perge2013,Nadj-perge2014}, or 3D time-reversal invariant (TRI) topological insulators (TI)\cite{Fu2008}. While much work has taken place on both groups of proposals, there have been few unambiguous signs of topological superconductivity observed experimentally. In this endeavor, the most promising experimental signatures have come from the second class of proposals, in particular the spin-orbit coupled semiconductors\cite{Oreg2010,Sau2010,Mourik2012,Marcus2016} and magnetic adatoms proximity-coupled with $s$-wave superconductors\cite{Nadj-Perge2013,Nadj-perge2014}. Nonetheless, it is clear that within each of the proposals to observe topological superconductivity, there is a trend in the components required to produce the unconventional superconductivity: non-zero Berry curvature induced by spin-orbit coupling and broken time reversal symmetry by magnetism.

Of the available platforms within which one may combine these ingredients, one of the well known routes to generate topological superconductivity is via the superconducting proximity effect in a heterostructure sample of a conventional $s$-wave superconductor and 3D TRI TI\cite{Fu2008}. In the pioneering work of Fu and Kane, Cooper pairs from $s$-wave superconductors that are proximity-coupled to 3D TRI TI tunnel from the superconductor into the TI resulting in the acquisition of an topological superconductivity that behaves an effective spinless, chiral $p_x+ip_y$ superconductor without breaking time-reversal symmetry. As compared to proposals using non-TI heterostructures\cite{Sau2010,Nadj-Perge2013} or intrinsic superconductors\cite{Jang2011,Kallin2009,Kallin2012,Hor2010a,Fu2010,Tafti2013}, the Fu-Kane proposal is attractive as it does not require further assumptions on any of the physical parameters such as Cooper pairing amplitudes between different orbitals\cite{Tanaka2010} or the position of the chemical potential\cite{Fu2010,Hung2013}.
To facilitate the generation of chiral edge states, a Zeeman field may be introduced to open a gap in the energy spectrum and thereby form a boundary at the surface of the TI\cite{Fu2008}.  
To this end, topological insulators with magnetic dopants that break time-reversal symmetry are of great interest\cite{Kulbachinskii2001,Dyck2002,Hor2010,Chen2010,Yu2010,Xu2012,Jiang2012,Jing2015,Jing2016} as a platform to observe topological superconductivity and chiral edge states. In this work, we study how introducing magnetic dopants affects the proximity induced superconductivity of the 3D TI system. Unlike a similar study that has been performed on magnetically-doped TI whose Zeeman field is randomly oriented within the TI\cite{Maciejko2016}, we consider magnetically ordered dopants through the addition of a uniform Zeeman splitting term in a thin 3D TRI TI sample such as Bi$_2$Se$_3$ to form a magnetic domain via the net exchange field\cite{Yu2010, Xu2012}. As experimental TI samples must be thin for superconductivity to be observed on the surface, we focus on the ``ultrathin'' limit of the TI where the surface states are not well-isolated but hybridized resulting in a gapped surface state spectrum.

%============FIGURE=============================
\begin{figure}[t!] 
  \centering
   \includegraphics[width=0.5\textwidth]{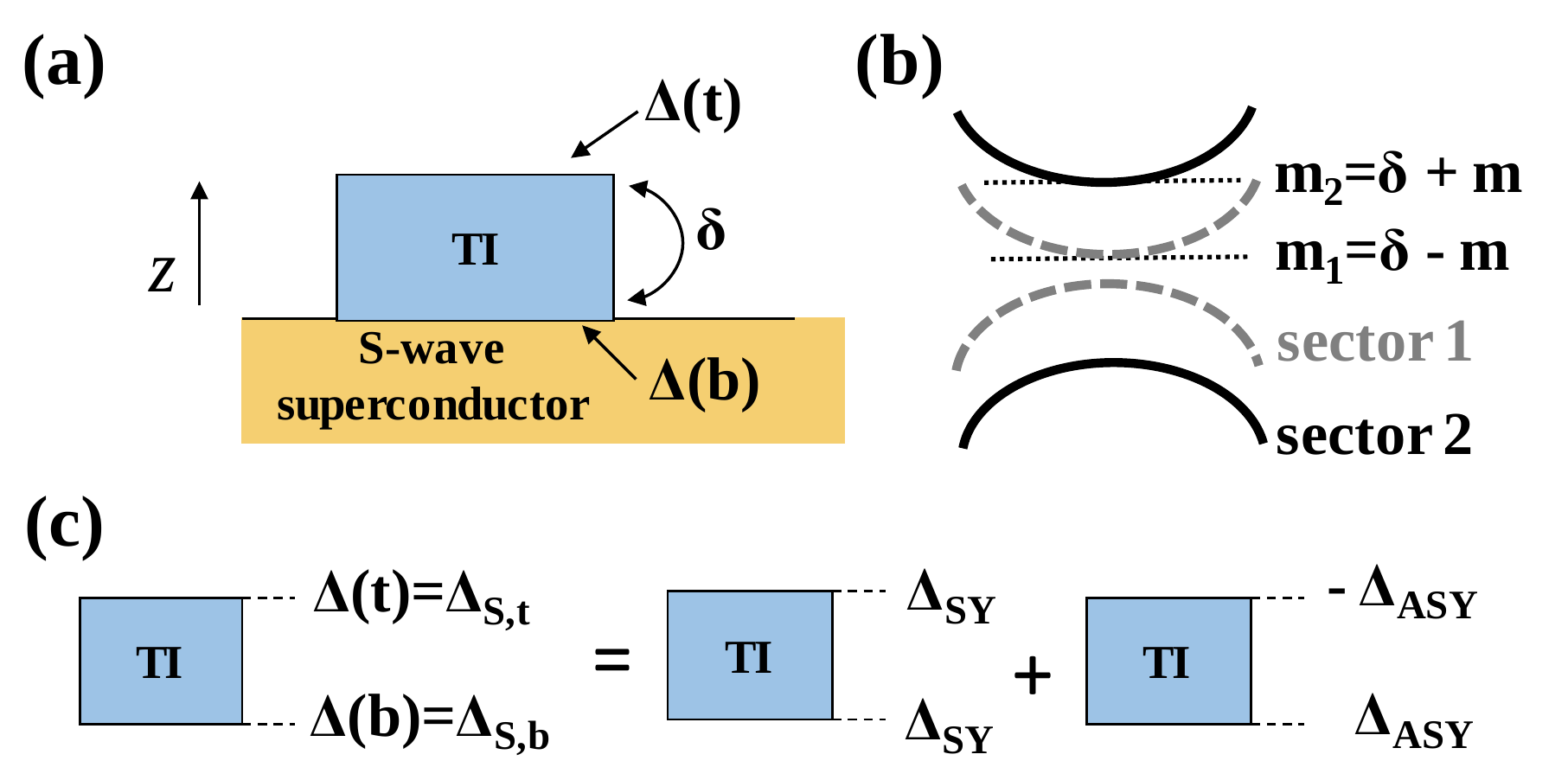}
  \caption{(a) A schematic of system under consideration. The ultrathin TI is grown on top of the s-wave superconductor, whose hybridization gap is described as $\delta$. The induced order parameter at top (bottom) surface is indicated as $\Delta(t)$ ($\Delta(b)$). (b) After we apply the proper rotation to the system, we obtain two decoupled systems in hybridization basis of the top and bottom surfaces. Two individual sectors are referred as to a sector 1 and sector 2. (c) Total induced superconducting order parameter is equivalent to a combination of (spatially) symmetric and anti-symmetric superconducting order parameter.}\label{fig:system}
\end{figure}   %==========FIGURE=================================

We seek to understand the physics of magnetically-doped, ultrathin TI and its topological phase by analyzing the superconducting order parameter using both analytical and numerical techniques. In Section ~\ref{sec:phenomenological}, we introduce a 2D continuum model for the surface states of an ultrathin TI that accounts for the hybridization gap. By applying a series of unitary transformations, we note that the model can be separated into independent sectors, whose relevant pairing potential have a symmetric and anti-symmetric spatial form when superconductivity is added. We then analyze the symmetric and anti-symmetric $s$-wave pairing potential at the phenomenological level by assuming a constant induced order parameter, and find that anti-symmetric pairing is dominant for experimentally relevant strengths of the Zeeman field. Simplifying the Hamiltonian with assumed anti-symmetric pairing potential, we find that gap closing points exist and are controlled by three parameters that can be tuned in experiment: the chemical potential, hybridization gap, and Zeeman energy. As our system is in $D$ class within the Altland-Zirnbauer classification\cite{Zirnbauer1996, Altland1997}, it is characterized by a $\mathbb{Z}$ topological invariant\cite{Ryu2008}. Thus we analyze the gap closing points and corresponding topological phase by evaluating the Chern number to obtain the resulting phase diagram.
In Section~\ref{sec:results3D}, we model a more realistic lattice system by self-consistently solving for the superconducting order parameter in a heterostructure of a $s$-wave superconductor coupled to a TI using the Bogoliubov-de Gennes (BdG) formalism. Our numerical simulation accurately captures the effects of bulk and surface bands that are present in TI and includes the spin dynamics of these bands when magnetism is introduced. An induced superconducting order parameter obtained from bulk states of the TI shows a rapid decay in magnitude with increasing magnetic impurity concentration as the Zeeman energy splits the band and suppresses the $s$-wave pairing. In contrast, the surface states show an induced order parameter that persists over an experimentally relevant range of the Zeeman energy due to their spin-momentum locked nature and non-zero projection of electron pairs into the $s$-wave pairing component. Moreover, self-consistent calculation shows that the anti-symmetric pairing potential is dominant at non-zero Zeeman energy, thereby, we confirm our phenomenological analysis.
Lastly, in Section~\ref{sec:conclusion}, we summarize our results and provide our concluding remarks.

\section{Induced superconductivity of ultra-thin TI at phenomenological level} \label{sec:phenomenological} 
 \subsection{Surface state model for ultra-thin TI} \label{sec:2Dmodel}

Fig. \ref{fig:system}(a) shows the system under consideration, in which a TI is placed on top of a conventional $s$-wave superconducting metal. We first consider the surface state Hamiltonian of TI, $H_{TI}=\sum_\vec{k} \Psi_\vec{k}^\dagger \hat{H}_{\text{TI}}(\vec{k}) \Psi_\vec{k}$, where $\Psi_{\vec{k}}=[c_{t,\uparrow}(\vec{k}),c_{t,\downarrow}(\vec{k}),c_{b,\uparrow}(\vec{k}),c_{b,\downarrow}(\vec{k})]^T$. We define $c_{t,\uparrow}(\vec{k})$ ($c_{b,\downarrow}(\vec{k})$) for electron annihilation operator at top (bottom) surface with up (down) spin states at $\vec{k}=(k_x,k_y)$. Then the Hamiltonian $\hat{H}_{\text{TI}}$ is defined as 
\begin{equation} \label{eq:Hsurf_bare}
		\hat{H}_{\text{TI}}(\vec{k})=
		\begin{pmatrix}
		\hat{H}_\text{top} & \delta I_2 \\
		\delta I_2 & \hat{H}_\text{bot} \\
		\end{pmatrix}-\mu, \\
\end{equation}
where $\sigma_{x,y,z}$ are the Pauli matrices for the spin degree of freedom, $I_2$ is a $2\times2$ identity matrix, and $2\delta$ is the hybridization gap. We use $\hat{H}_\text{top}=-k_y\sigma_x+k_x\sigma_y$ and $\hat{H}_\text{bot}=k_y\sigma_x-k_x\sigma_y$ for the low energy description of top and bottom surface state Hamiltonians, respectively\cite{Zhang2010}, to form a hybridized Hamiltonian that preserves time reversal symmetry. 
As the focus of this work is to understand the influence of magnetic dopants, we add the effect of a uniform perpendicular magnetization to the proximity-coupled TI surface via addition of the Zeeman energy splitting term within the Hamiltonian in Eq.~(\ref{eq:Hsurf_bare}). The Zeeman term whose principle axis is aligned in $\hat{z}$ takes the form $\hat{H}_{\text{Zeeman}}=(m_z)I_2\otimes \sigma_z$, where $m_z\sigma_z$ is the energetic splitting of spin states due to the magnetization arising from the dopants included within the TI film, and the identity matrix acts on top and bottom surface (or pseudospin) degree of freedom. Previous work considering the effects of magnetic dopants on the surface physics of TRI TI indicates that the addition of magnetically ordered impurities allows for the development of a net ferromagnetic order,\cite{Liu2009,Rosenberg2012,Qin2014}  and thus a uniform magnetization can provide a simple but accurate picture of the TI with magnetic dopants. Within this work, we ignore any orbital effects  as those resulting from magnetic dopants are negligible\cite{Thonhauser2011}.

Including the Zeeman term, we may write Eq. (\ref{eq:Hsurf_bare}) in $4\times 4$ matrix form as
\begin{equation} \label{eq:Hsurf}
\begin{split}
&\hat{H}_{\text{surf}}(\vec{k})=\hat{H}_{\text{TI}}(\vec{k})+\hat{H}_{\text{Zeeman}}= \\
		&
		\begin{pmatrix}
		m_z-\mu & -k_y-ik_x & \delta & 0 \\
		-k_y+ik_x & -m_z-\mu & 0 & \delta \\
		\delta & 0 & m_z-\mu & k_y+ik_x \\
		0 & \delta & k_y-ik_x & -m_z-\mu \\
		\end{pmatrix}. \\
\end{split}
\end{equation}
Eq. (\ref{eq:Hsurf}) is further simplified by applying proper rotational matrices. Without loss of generality, we remove $k_y$ by applying an $SU(2)$ rotation, $U_1=I_2\otimes e^{i\theta_1\sigma_z}$, where $\theta_1=-\tan^{-1}(k_x/k_y)/2$. In addition, we apply another $SU(2)$ rotation on the pseudo-spin degree of freedom (top/bottom layer), $U_2=e^{i\frac{\pi}{4}\tau_y}\otimes I_2$, where $\tau_y$ is the Pauli matrix for the pseudospin degree of freedom, and rewrite the Hamiltonian in Eq. (\ref{eq:Hsurf}) using a hybridized basis of top and bottom layer. The resultant Hamiltonian is
\begin{equation} \label{eq:Hsurf2}
\hat{H}'_{\text{surf}}(k)=U_2U_1\hat{H}_{\text{surf}}(k_x,k_y)U_1^\dagger U_2^\dagger .
\end{equation}
We may write Eq. (\ref{eq:Hsurf2}) in $4\times 4$ matrix form:
\begin{equation} \label{eq:Hsurf22}
\hat{H}'_{\text{surf}}(k)=
\begin{pmatrix}
m_2-\mu & 0 & 0 & -k \\
0 & m_1-\mu & -k & 0 \\
0 & -k & -m_1-\mu & 0 \\
-k & 0 & 0 & -m_2-\mu \\
\end{pmatrix},
\end{equation}
where $k=|\vec{k}|$. In Eq. (\ref{eq:Hsurf22}), we define effective Zeeman energy $m_1=\delta-m_z$ and $m_2=\delta+m_z$.
Note that the Hamiltonian Eq. (\ref{eq:Hsurf22}) is now decoupled into two sectors:
\begin{equation} \label{eq:Hsurf12}
\begin{split}
\hat{H}'_{\text{surf,1}}(k)=&m_1\tau_z-k\tau_x-\mu, \\
\hat{H}'_{\text{surf,2}}(k)=&m_2\tau_z-k\tau_x-\mu, \\
\end{split}
\end{equation}
where $\tau_{x,y,z}$ is the Pauli matrix in pseudospin space whose basis is now in linear combination of spin and layer degree of freedoms.
For the following arguments, $\hat{H}_{\text{surf,1}}$ and $\hat{H}_{\text{surf,2}}$ refers to the Hamiltonian in sector 1 and sector 2, respectively, as shown in Fig. \ref{fig:system} (b).
The rotated basis for Eq. (\ref{eq:Hsurf2}) is
\begin{equation} \label{eq:U2c}
U_2
\begin{pmatrix}
c_{t,\uparrow} \\ c_{t,\downarrow} \\ c_{b,\uparrow} \\ c_{b,\downarrow} \\
\end{pmatrix}
=\frac{1}{\sqrt{2}}
\begin{pmatrix}
c_{t,\uparrow}+c_{b,\uparrow} \\ c_{t,\downarrow}+c_{b,\downarrow} 
\\ -c_{t,\uparrow}+c_{b,\uparrow} \\  -c_{t,\downarrow}+c_{b,\downarrow}  \\
\end{pmatrix}
=
\begin{pmatrix}
\nu_{2,\uparrow} \\ \nu_{1,\downarrow} \\ 
\nu_{1,\uparrow} \\  \nu_{2,\downarrow} \\
\end{pmatrix},
\end{equation}
where we omit in-plane momentum index $\vec{k}$ for simplicity. Eq. (\ref{eq:U2c}) shows that the electron states at top and bottom surface form bonding and anti-bonding like hybridized states, whose annihilation operator is defined by $\nu$. Due to the opposite helicity of top and bottom surfaces, the hybridized basis satisfies following unitary transformation 
\begin{equation} \label{eq:helicity}
(\tau_x\otimes\sigma_z)^\dagger \hat{H}_{\text{surf}}(\tau_x\otimes\sigma_z)=\hat{H}_{\text{surf}}, 
\end{equation}
which exchanges the particles at top and bottom surfaces, namely, $c_{t,\uparrow}\rightarrow c_{b,\uparrow}$ and $c_{t,\downarrow}\rightarrow -c_{b,\downarrow}$.
Consequently, a band having a basis of $\nu_{1,\downarrow}=(c_{t,\downarrow}+c_{b,\downarrow})/\sqrt{2}$ has $\nu_{1,\uparrow}=(-c_{t,\uparrow}+c_{b,\uparrow})/\sqrt{2}$ as another basis to satisfy Eq. (\ref{eq:helicity}) and this particular combination of basis forms the bands in sector 1. The other set of basis states forms the bands in sector 2. Having two well separated sectors in our Hamiltonian, we may analyze possible superconductor pairing order parameters.

\subsection{Symmetric and anti-symmetric superconducting order parameter} \label{sec:2DOP}

When a surface state is coupled to a superconducting system, one may observe an induced superconducting order parameter in the surface states of the TI as Cooper pairs tunnel from the superconductor system\cite{Klapwijk1995}. In our system, the induced order parameters in the top and bottom surface differ in their magnitude as the Cooper pair tunneling probability decays as a function of a spatial separation from the interface between TI and $s$-wave superconductor to the other surface of the TI\cite{Xu2014}. Due to this gradient in order parameter magnitude, we may decompose the induced order parameter into two distinct components, whose individual $U(1)$ phases are symmetric and anti-symmetric in $\hat{z}$ direction. Fig. \ref{fig:system}(c) shows the simplest example, e.g. $\Delta(z)=\Delta_{SY}(z)+\Delta_{ASY}(z)$ where $\Delta_{SY}(t)=\Delta_{SY}(b)$ and $\Delta_{ASY}(t)=-\Delta_{ASY}(b)$. 

Having the basis representation in Eq. (\ref{eq:U2c}), we may examine the induced pair correlation function defined within each of the sectors or across different sectors. A pair correlation function of the intra-band $s$-wave pairing (within sector 1 or sector 2) is, for example, 
\begin{equation} \label{eq:intra}
F_\text{intra}=\langle v_{1,\uparrow} v_{1,\downarrow}\rangle
=\frac{1}{2}\langle (-c_{t,\uparrow}+c_{b,\uparrow})(c_{t,\downarrow}+c_{b,\downarrow}) \rangle,
\end{equation}
where we suppress the in-plane momentum index $\vec{k}$ for brevity.
Eq. (\ref{eq:intra}) shows an odd parity under the exchange of the layer degree of freedom, or,  $F_\text{intra} \xrightarrow{ t\leftrightarrow b} -F_\text{intra}$. By defining the on-site $s$-wave order parameter\cite{Tinkham1996} as $\Delta_\text{intra}=\sum_\vec{k} F_\text{intra}(\vec{k})$, we find that the intra-band $s$-wave order parameter is anti-symmetric in the $\hat{z}$ direction.
Meanwhile, a pair correlation function of the inter-band $s$-wave pairing (between sector 1 and sector 2) is, for example,
\begin{equation}  \label{eq:inter}
F_\text{inter}=\langle v_{1,\uparrow} v_{2,\downarrow}\rangle
=\frac{1}{2}\langle (-c_{t,\uparrow}+c_{b,\uparrow})(-c_{t,\downarrow}+c_{b,\downarrow}) \rangle,
\end{equation}
which shows an even parity under the exchange of the layer degree of freedom, or, $F_\text{inter} \xrightarrow{ t\leftrightarrow b} +F_\text{inter}$ and, thus, the resultant inter-band $s$-wave order parameter, $\Delta_\text{inter}=\sum_\vec{k} F_\text{inter}(\vec{k})$, is symmetric in $\hat{z}$ direction. 
Therefore, Eq. (\ref{eq:intra}) shows that the anti-symmetric order parameter is responsible for superconducting pairing that occurs within the same sectors, whereas Eq. (\ref{eq:inter}) shows that the symmetric order parameter results in superconducting pairing across different sectors.

The above argument is explicitly shown by constructing the Bogoliubov-de Gennes (BdG) Hamiltonian. We add an induced order parameter to the system as a constant value, $\Delta$, at the phenomenological level\cite{Fu2008}. Then the BdG Hamiltonian is
\begin{equation} \label{eq:BdG}
\hat{H}_{\text{BdG}}(\vec{k})=
\begin{pmatrix}
\hat{H}_{\text{surf}}(\vec{k}) & \hat{H}_{\text{BCS}} \\
\hat{H}_{\text{BCS}}^\dagger & -\hat{H}_{\text{surf}}^*(-\vec{k}) \\
\end{pmatrix},
\end{equation} 
where the $s$-wave pairing Hamiltonian is
\begin{equation} \label{eq:Hpair}
\hat{H}_{\text{BCS}}=
\begin{pmatrix}
\Delta_{t} & 0 \\
0 & \Delta_{b} \\
\end{pmatrix}
\otimes i\sigma_y,
\end{equation}
where $\Delta_{t}$ and $\Delta_{b}$ are the $s$-wave order parameter at top and bottom surface, respectively.
We then decompose the order parameter in Eq. (\ref{eq:Hpair}) into two components, namely, the symmetric and anti-symmetric components. Fig. \ref{fig:system}(c) depicts symmetric and anti-symmetric components of superconducting order parameter written as
\begin{align} 
\hat{H}_{\text{BCS,SY}}=
\begin{pmatrix}
\Delta_{SY} & 0 \\
0 & \Delta_{SY} \\
\end{pmatrix}\otimes i\sigma_y, \label{eq:BCSSY} \\
\hat{H}_{\text{BCS,ASY}}=
\begin{pmatrix}
-\Delta_{ASY} & 0 \\
0 & \Delta_{ASY} \\
\end{pmatrix}\otimes i\sigma_y \label{eq:BCSASY},
\end{align}
respectively, where $\Delta_{SY}$ and $\Delta_{ASY}$ are corresponding decomposed order parameters.
By defining $\bar{U}_{1,2}=\begin{pmatrix}U_{1,2}& 0 \\ 0 & U_{1,2}^*\end{pmatrix}$, we obtain a simplified Hamiltonian in BdG form, 
\begin{equation} \label{eq:BdGp}
\begin{split}
\hat{H}_{\text{BdG}}'(k)=&\bar{U}_2\bar{U}_1 \hat{H}_{\text{BdG}}(\vec{k})\bar{U}_1^\dagger \bar{U}_2^\dagger  \\
=&
\begin{pmatrix}
\hat{H}_{\text{surf}}'(k) & \hat{H}'_{\text{BCS,SY}}+\hat{H}'_{\text{BCS,ASY}} \\
h.c. & -\hat{H}_{\text{surf}}'(-k) \\
\end{pmatrix},
\end{split}
\end{equation}
where the symmetric and anti-symmetric component of pairing Hamiltonian becomes
\begin{equation} \label{eq:sym}
 \hat{H}'_{\text{BCS,SY}} =
 \begin{pmatrix} 
 0 & \Delta_{SY} & 0 & 0 \\
 -\Delta_{SY} & 0 & 0 & 0 \\
 0 & 0 & 0 & \Delta_{SY} \\
 0 & 0 & -\Delta_{SY} & 0 \\
 \end{pmatrix},
\end{equation}
and
\begin{equation} \label{eq:asym}
 \hat{H}'_{\text{BCS,ASY}} =
 \begin{pmatrix}
 0 & 0 & 0 & \Delta_{ASY} \\
0 & 0 & -\Delta_{ASY} & 0 \\
 0 & \Delta_{ASY} & 0 & 0 \\
 -\Delta_{ASY} & 0 & 0 & 0 \\
 \end{pmatrix},
\end{equation}
respectively.
The pairing Hamiltonian in Eq. (\ref{eq:sym}) explicitly shows that an electron (hole) band in sector 1 is coupled with the hole (electron) band in sector 2, which we refer to as an inter-band pairing. Likewise, Eq. (\ref{eq:asym}) shows that the anti-symmetric superconducting order parameter couples electron bands and hole bands in the same sector, which we refer to as intra-band pairing.

\subsection{Symmetric and anti-symmetric superconducting order parameter at non-zero Zeeman energy} \label{sec:2DOP2}

Although the hybridized surface state bands are initially degenerate having both the symmetric and anti-symmetric superconducting pairing, introducing the Zeeman term splits them in energy and changes their relative contributions to the resultant superconductivity. 
In order to gain insight on the superconducting order parameter evolution as a function of the Zeeman energy, we examine pair correlation function of the TI surface states.
We start with the BdG Hamiltonian defined by
\begin{equation} \label{eq:Htot}
\tilde{H}_{TI+SC}=
\begin{pmatrix}
\tilde{H}_{top} & \tilde{\delta} & 0 \\
\tilde{\delta}^\dagger & \tilde{H}_{bot} & \tilde{t}_c \\
0 & \tilde{t}_c^\dagger & \tilde{H}_{SC} \\
\end{pmatrix},
\end{equation}
where $\tilde{H}_{top}$, $\tilde{H}_{bot}$ are top and bottom surface TI Hamiltonian in BdG form, respectively. The top and bottom surface are connected by the hopping matrix $\tilde{\delta}$, $\tilde{H}_{SC}$ is $s$-wave superconductor Hamiltonian connected to the bottom surface of the TI by the hopping matrix $\tilde{t}_c$. The form of the hopping matrices $\tilde{\delta}$ and $\tilde{t}_c$ are general and their specific definition is dependent on specific basis chosen.
Then Eq. (\ref{eq:Htot}) satisfies $(\hat{H}_{TI+SC}-\epsilon)\Psi_{TI+SC}(\vec{r})=0$, where $\Psi_{TI+SC}(\vec{r})$ is the wavefunction at $\vec{r}=(x,y)$ and energy $\epsilon$. By integrating out the superconductor degree of freedom, and assuming that tunneling from $s$-wave superconductor to bottom surface of the TI is local in space\cite{Stanescu2010, Sau2010b}, we now have
\begin{equation} \label{eq:Htot2}
\tilde{H}_{TI}=
\begin{pmatrix}
\tilde{H}_{top} & \tilde{\delta}  \\
\tilde{\delta}^\dagger & \tilde{H}_{bot}+\tilde{\Sigma}_{SC} \\
\end{pmatrix},
\end{equation}
which satisfies $(\tilde{H}_{TI}-\omega)\Psi_{TI}(\vec{r})=0$ where $\Psi_{TI}(\vec{r})$ is the wavefunction of TI system. In Eq. (\ref{eq:Htot2}), the proximity-induced superconductivity comes into play through the $s$-wave superconductor self-energy term, $\tilde{\Sigma}_{SC}=-\tilde{\delta}\tilde{G}_{SC}\tilde{\delta}^\dagger$, where $\tilde{G}_{SC}=(\tilde{H}_{SC}-\epsilon)^{-1}$ is the superconductor Green's function\cite{Sau2010b}. 
Specifically, we only consider the off-diagonal part (or the anomalous part of the Green's function) of the self-energy term to elucidate qualitative behavior of the pair correlation function. By adopting the energy independent self-energy term near $\epsilon\sim0$, we find a simple expression,\cite{Stanescu2010, Sau2010b} $\tilde{\Sigma}_{SC}=\Delta_0i\sigma_y$, for the $s$-wave superconductor self-energy term. 

To utilize our analysis in section \ref{sec:2Dmodel} and \ref{sec:2DOP}, we adopt the low energy description of the ultrathin TI surface states Hamiltonian in Eq. (\ref{eq:Hsurf}) and its BdG form in Eq. (\ref{eq:BdG}). We set order parameter in Eq. (\ref{eq:Hpair}) as $\Delta_{t}=0$ and $\Delta_{b}=\Delta_0$ following the discussion in Eq. (\ref{eq:Htot2}).
Then we evaluate the top (t) and bottom (b) surface on-site $s$-wave pair correlation function,
\begin{equation} \label{eq:Deltastb}
\begin{split}
F_{t/b}=&\sum_{\vec{r}}\braket{c_{t/b,\uparrow}(\vec{r})c_{t/b,\downarrow}(\vec{r})
-c_{t/b,\downarrow}(\vec{r})c_{t/b,\uparrow}(\vec{r})} \\
=& \sum_{\vec{k}}
[F_{t/b}^{\uparrow\downarrow}(\vec{k}) - F_{t/b}^{\downarrow\uparrow}(\vec{k})], \\
\end{split}
\end{equation}
which explicitly captures the superconducting proximity effect\cite{Black2011}. In Eq. (\ref{eq:Deltastb}), we define momentum space resolved correlation function, $F_{t/b}^{\uparrow\downarrow}(\vec{k})=\braket{c_{t/b,\uparrow}(\vec{k})c_{t/b,\downarrow}(-\vec{k})}$. 
For detailed calculation for the correlation function, see Appendix \ref{app:op}. In this work, we are interested in the coupled system where the surface states of the TI and metallic states in superconductor are not strongly mixed, which is a valid regime for realistic proximity-coupled TI systems\cite{Kun2011,Hassan2014,Xu2014}. Then we may assume that the host superconductor's order parameter is unaffected by the TI and, as a result, we compute the pair correlation function $F_{t/b}$ without considering self-consistency. 
We now examine spatial distribution of the order parameter by defining the symmetric and anti-symmetric pair correlation functions, 
\begin{equation} \label{eq:DeltaSYASY}
F_{SY}=(F_{b}+F_{t})/2, \;\;
F_{ASY}=(F_{b}-F_{t})/2.
\end{equation}
%============FIGURE=============================
\begin{figure}[t!]  
  \centering
   \includegraphics[width=0.45\textwidth]{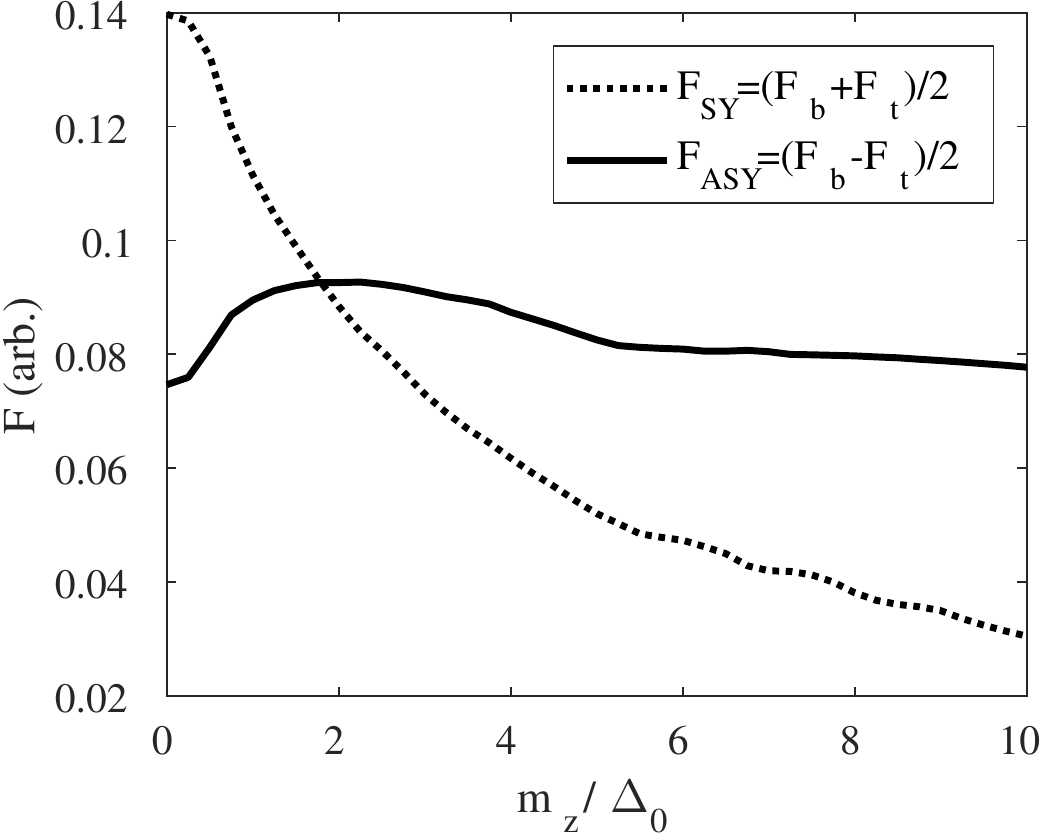}
  \caption{Symmetric and anti-symmetric pair correlation function of the top and bottom surfaces of the TI. The hybridization gap of the system is set to $\delta=0.5$ eV. We set the chemical potential of TI $\mu/\delta=1.5$, whose location crosses the hybridized surface band. We set $\Delta_0/\delta=0.1$, and cut-off energy is set to be $|E_c/\delta|=1.0$.}\label{fig:FSY}
\end{figure}   %==========FIGURE=================================
In Fig. \ref{fig:FSY}, we show the symmetric and anti-symmetric pair correlation function $F_{SY}$ and $F_{ASY}$ as a function of the Zeeman energy. Although, based on Fig. \ref{fig:FSY}, we may conclude that the symmetric pairing potential determines the system order parameter at the zero or low Zeeman energy, the pair correlation function becomes dominantly anti-symmetric as we increase the Zeeman energy.
This particular transition may be understood in terms of the inter-band and intra-band pairings described in Eq. (\ref{eq:sym}) and (\ref{eq:asym}), which are governed by the symmetric and anti-symmetric pairing potential, respectively.
%Following the discussion in section \ref{sec:2Dmodel}, the ultrathin TI system is described by two decoupled subspaces, namely, sector 1 and 2 in the presence of the Zeeman energy. As a result, the pairing within the same sector (intra-band pairing) and across sectors (inter-band pairing) evloves differently as a function of the Zeeman energy.  

\emph{Inter-band pairing}: The ultrathin TI system is described by two decoupled subspaces, namely, sector 1 and sector 2, as shown in Eq. (\ref{eq:Hsurf22}) in hybridized basis.
The non-zero Zeeman energy splits sector 1 and sector 2 in energy and, 
%the splitting results in a Fermi surface mismatch between two sectors. 
for example, electron states at Fermi wavevector $\vec{k}_F$ in sector 1 no longer has a pair at $-\vec{k}_F$ in sector 2.
Consequently, the Cooper pairs formed by the inter-band pairing experience a Fermi surface mismatch as the Zeeman energy splits the bands in energy. Further increase in the Zeeman energy results in a larger mismatch in Fermi surface that costs more energy to form Cooper pairs reducing the number of pairing states. For ferromagnetically doped superconducotor, for instance, the system eventually undergoes a phase transition from superconducting to normal when the Zeeman energy is as large as the superconducting gap at zero field and zero temperature\cite{Clogston1962, Sarma1963, Izuyama1969}. Although we may not see such phase transition in proximity coupled system, BCS formalism captures this Fermi surface mismatch. As a result of the Fermi surface mismatch, Fig. \ref{fig:FSY} shows a rapid decrease of the symmetric pair correlation function (dashed line). 

\emph{Intra-band pairing}: In the absence of the Zeeman energy, one can observe $s$-wave electron pairing within each band due to the spin-momentum locked nature of the surface states. In the presence of the Zeeman field, however, the TI surface states exhibit canted ``hedgehog'' spin texture\cite{Xu2012} and, consequently, the projected $s$-wave pairing magnitude decreases due to the out-of-plane canting induced by the Zeeman field. However, each band still possesses non-zero $s$-wave pairing as the paired electrons experience no Fermi surface mismatch, unlike those in the inter-band pairing case. As a result, Fig. \ref{fig:FSY} shows the anti-symmetric pair correlation function (solid line) persisting with increasing Zeeman field.

The above analysis shows that the anti-symmetric pairing potential is the dominant factor to form induced superconductivity for Zeeman energies whose scale is larger than the zero-field superconducting gap. This is the experimentally relevant regime considering that the typical superconducting gap is few meV for 4-quintuple-layer Bi$_2$Se$_3$\cite{Xu2014} while the Zeeman energy may be as large as $\sim 50$ meV in $2.5$\% Mn doped Bi$_2$Se$_3$\cite{Xu2012} system. In this regard, the anti-symmetric pairing potential is more relevant than the symmetric one for magnetically-doped ultrathin TI system.

\subsection{Topologically non-trivial phases in ultrathin TI} \label{sec:phase}

Our analysis in Section \ref{sec:2DOP} shows that the anti-symmetric pairing potential plays a major role in superconductivity in the magnetically-doped ultrathin TI system when the induced Zeeman energy is larger than the zero-field induced superconducting gap. Taking advantage of the property of the anti-symmetric pairing potential discussed in Eq. (\ref{eq:asym}), the BdG Hamiltonian in Eq. (\ref{eq:BdGp}) is now decoupled into two independent sectors. With the reduced BdG Hamiltonian, we now obtain the analytic form of the quasi-particle spectrum and use the gap closing points to identify topologically non-trivial phases. 

From Eqs. (\ref{eq:Hsurf22}, \ref{eq:BdGp}, \ref{eq:asym}), we obtain a block diagonal form of the BdG Hamiltonian, which we rewrite as two decoupled BdG Hamiltonian for sector 1 and sector 2, 
\begin{equation} \label{eq:BdGasym}
\hat{H}'_{\text{BdG,i}}=
\begin{pmatrix}
\hat{H}'_{\text{surf,i}}(k) & -\Delta_0 i\tau_y \\
-(\Delta_0 i\tau_y)^\dagger  & -{\hat{H'}}^*_{\text{surf,i}}(-k) \\
\end{pmatrix}.
\end{equation}
In Eq. (\ref{eq:BdGasym}), $\tau_y$ is the Pauli matrix whose representation is in the hybridized basis shown in Eq. (\ref{eq:U2c}), and $\hat{H}'_{\text{surf,i}}(k)$ is the sector $i=(1,2)$ Hamiltonian in Eq. (\ref{eq:Hsurf12}). From this, we obtain the quasi-particle spectrum
\begin{equation} \label{eq:HBdGasymE}
E_{i}(k) = \pm\sqrt{k^2+m_{i}^2+\Delta_0^2+\mu^2 
\pm 2\sqrt{m_{i}^2\Delta_0^2 + k^2\mu^2 + m_{i}^2\mu^2} },
\end{equation}
where $m_{1}=\delta-m_z$ and $m_2=\delta+m_z$ is the effective magnetic field defined in Eq. (\ref{eq:Hsurf22}) for each of the sectors. For non-zero $\Delta_0$, we find that a gap closing occurs only at $k=0$ when $|m_i|=\sqrt{\Delta_0^2+\mu^2}$. The system undergoes a phase transition at the identified gap closing point\cite{Murakami2007} with the Zeeman energy $m_z=\delta+\sqrt{\Delta_0^2+\mu^2}$ and, therefore, the gap closing point is determined by the system parameters: the hybridization gap ($\delta$), the position of the chemical potential ($\mu$), and the induced superconducting gap ($\Delta_0$). As a topological classification of our system is in $D$ class\cite{Ryu2008}, the relevant phase is classified by the $\mathbb{Z}$ topological invariant. In this regard, we numerically evaluate the Chern number by varying the chemical potential and Zeeman energy, whose values are normalized by the hybridization gap. 

Prior to our analysis on the system Chern number, we first define a lattice regularized model for the top surface Hamiltonian\cite{Zhang2010}
\begin{equation} \label{eq:H2D}
\begin{split}
\hat{H}_{\text{top}}^{\text{latt}}=&(\hbar v_F/a)[-\sin (k_ya)\sigma_x+\sin (k_xa)\sigma_y] \\
& +(D/a^2)[2-\cos (k_xa)-\cos (k_ya)]\sigma_z,
\end{split}
\end{equation}
and bottom surface Hamiltonian is $\hat{H}_{\text{bot}}^{\text{latt}}=-\hat{H}_{\text{top}}^{\text{latt}}$, where $a$ is a lattice constant, $v_F$ is the Fermi velocity, $D$ is a parameter that controls the quadratic term at higher energy. Without loss of generality, we set $a=1$ and $\hbar v_F=D=1$. The second line of the Eq. (\ref{eq:H2D}) gaps out the extra Dirac points at $(0,\pi),\;(\pi,0),\;(\pi,\pi)$. Although this term breaks time-reversal symmetry, the system under consideration has broken time-reversal symmetry due to the magnetic dopants. Therefore, Eq. (\ref{eq:H2D}) correctly captures relevant low energy physics in our system. 
Using the Eqs. (\ref{eq:BdG}, \ref{eq:H2D}), we construct the BdG Hamiltonian to obtain eigenstates of the $n$th band to calculate corresponding Chern number\cite{Hatsugai2005}, $\tilde{c}_n$. For calculation details of the Chern number, see Appendix \ref{app:Chern}. Finally, the total Chern number of the system is obtained using $\tilde{c}=\sum_n \tilde{c}_n$, where $n$ runs over all of the occupied bands.

We return to the discussion below Eq. (\ref{eq:BdGasym}) to examine the Chern number of the BdG Hamiltonian for the sector 1 with the assumed anti-symmetric pairing potential. Assuming the hybridization gap satisfies $\delta>\sqrt{\Delta_0^2+\mu^2}$, the proximity coupled ultrathin TI system is initially gapped and in the trivial regime for $m_z<\delta-\sqrt{\Delta_0^2+\mu^2}$. As we increase the magnetic dopant concentration and, as a result, the Zeeman energy $m_z$, the Chern number of the system becomes -1 after the first gap closing point at $m_z=\delta-\sqrt{\Delta_0^2+\mu^2}$ and becomes -2 after the second gap closing point at $m_z=\delta+\sqrt{\Delta_0^2+\mu^2}$. Other gap closing points exist at $\vec{k}=(\pi,0)$ and $(0,\pi)$ for $|m_z-\delta|=2D/a^2\pm\sqrt{\Delta_0^2+\mu^2}$ and at $\vec{k}=(\pi,\pi)$ for $|m_z-\delta|=4D/a^2\pm\sqrt{\Delta_0^2+\mu^2}$. As a result, the Chern number becomes $0$ in the large $m_z$ limit. However, we only consider low-energy physics in this work, thus we only consider the gap closing point at $\vec{k}=(0,0)$. A similar analysis may be applied for $m<0$ using the Hamiltonian in sector 2. In fact, the Hamiltonian in Eq. (\ref{eq:BdGasym}) is analogous to the surface states of the TI system proximity coupled with the domain of the $s$-wave superconductor and ferromagnet (3D TI-SC-FM)\cite{Fu2008,Nagaosa2009,Andrei2013}. The analogy is made evident by taking the limit of $\delta\rightarrow0$, where the corresponding gap closing point converges to that of the 3D TI-SC-FM system. For example, by setting $\mu=0$, the gap closing points are at  $m_z=\delta\pm\sqrt{\Delta_o^2+\mu^2}\rightarrow\pm\Delta_0$, which coincide with the phase transition points of the 3D TI-SC-FM system\cite{Andrei2013}.
%
%============FIGURE=============================
 \begin{figure}[t!] 
  \centering
   \includegraphics[width=0.5\textwidth]{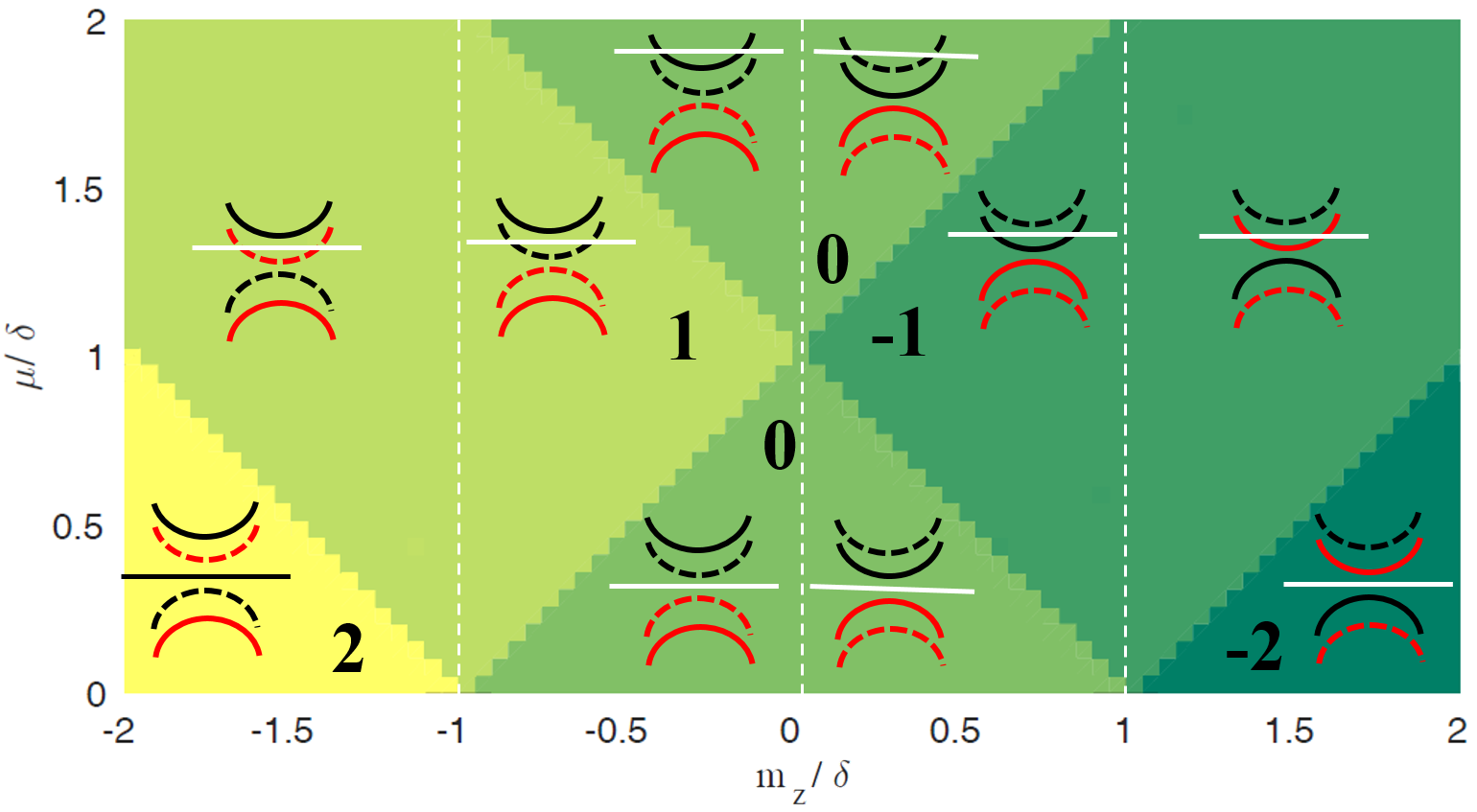}
  \caption{The phase diagram and corresponding Chern number of an ultrathin TI as a function of the chemical potential, $\mu$, and the Zeeman energy $m$. Both axises are normalized by the hybridization gap, $\delta$. For illustration purposes, we draw schematics that describe the motion of the surface state bands in different sectors and the corresponding Chern numbers. Note that solid curve (dashed curve) represents surface state band in sector 1 (sector 2), and black (red) color represents conduction (valence) band when $m=0$. The horizontal line in each band diagram corresponds to the position of the chemical potential. The $s$-wave order parameter at top and bottom surface is $\Delta_0=0$ and $0.01\delta$, respectively.}\label{fig:phase}
\end{figure}   %==========FIGURE=================================

To obtain a generic phase diagram, we assume that both the symmetric and anti-symmetric pairing potential are present. Specifically, we assume that the magnitude of the induced superconducting gap satisfies the experimentally relevant regime for the ultrathin TI system\cite{Xu2014}, namely, $\Delta_0\ll\delta$. In this regime, the specific form of the pairing potential is insignificant as we construct the phase diagram with other parameters relative to $\delta$. Thus, we set the pairing Hamiltonian to satisfy $\Delta_{t}=0,\, \Delta_{b}=\Delta_0$. 
Then we compute the Chern number and construct the phase diagram of the ultrathin TI in Fig. \ref{fig:phase} by varying the chemical potential, $\mu$, and the Zeeman energy, $m_z$, normalized by the hybridization gap, $\delta$. 
Each region in the phase diagram is illustrated by a schematic of the surface state band diagram in a normal phase TI with a specified location of the chemical potential.
Specifically, solid and dashed curves are the bands in sector 1 and sector 2, respectively, and the horizontal line indicates a location of the chemical potential. Note that the normal phase of the ultrathin TI undergoes the quantum phase transition\cite{Zyuzin2011} at $|m_z|=\delta$ due to the crossing of the conduction and valance band, which are initially separated by the hybridization gap. In order to capture this quantum phase transition point in our schematics, the initial valence and conduction bands at $m_z=0$ are indicated by red and black color, respectively.
To illustrate the system behavior in more detail, we follow, for example, the $\mu/\delta=0.8$ cut in Fig. \ref{fig:phase}. At $m_z/\delta=0$, the chemical potential is located within the hybridization gap and the system is in trivial phase (the Chern number is 0). As we increase the Zeeman energy, the conduction band is shifted and touches the chemical potential at $k=0$, which corresponds to a gap closing point in the quasi-particle spectrum of BdG Hamiltonian. Then, the band acquires a non-trivial Chern number of -1. Note that the valence band and conduction band are shifted further due to the Zeeman energy until they touch at $m_z/\delta=1$. However, this particular band crossing of the conduction and valence band has no effect on the Chern number, as there is no gap closing in quasi-particle spectrum of the BdG Hamiltonian. Once the Zeeman energy passes the second gap closing point, the additional band crossing at the chemical potential results in a Chern number of -2. Therefore, it is important to place the relevant parameters such as chemical potential and Zeeman energy in the right position to observe the topological superconductivity in the system. Further increase in Zeeman energy, $m_z$, will result in a gap closing at other high symmetry points, such as $\vec{k}=(\pi,0)$ and $(\pi,\pi)$, which eventually results in the Chern number becoming $0$ in the $|m_z|\rightarrow \infty$ limit. However, this is not important as we present the phase diagram within the experimentally relevant region where the Zeeman energy of the magnetic dopants is comparable to the hybridization gap.  

The physics of the non-trivial phase in the system may be further described by mapping our system to one of the well known topological superconducting systems. The Hamiltonian in Eq. (\ref{eq:BdGasym}) has a close analogy to the surface states of the 3D TI (for example, a top surface) proximity coupled to the interface of a ferromagnet and $s$-wave superconductor\cite{Andrei2013, Fu2008, Nagaosa2009}, where one may find chiral Majorana edge modes at the domain wall of the ferromagnet and superconductor region. Likewise, by properly choosing the concentration of the magnetic dopants and chemical potential of the ultrathin TI, we may also find the chiral edge states at the domain wall of the magnetically doped and undoped region of our system.

\section{Induced superconductivity of ultra-thin TI beyond phenomenological level} \label{sec:results3D} 
Our motivation in this section is to examine the induced order parameter beyond the phenomenological model presented in Section \ref{sec:phenomenological}. Instead of assuming the induced order parameter is a constant value, we examine the self consistently calculated induced order parameter by adopting a four band 3D TI model coupled to an $s$-wave superconductor.

\subsection{Effective model for 3D TI}  \label{sec:3DTI}

%%%%%%%%%%%%%%%%%%%%%%%%%%%%%%%%%%%%%%%%%%%%%%%%%%%%%%%%%%%%%%%%%%
%%%%    Schematic (Figure 1)
%%%%%%%%%%%%%%%%%%%%%%%%%%%%%%%%%%%%%%%%%%%%%%%%%%%%%%%%%%%%%%%%%%
\begin{figure}[t]
\includegraphics[width=0.8\columnwidth]{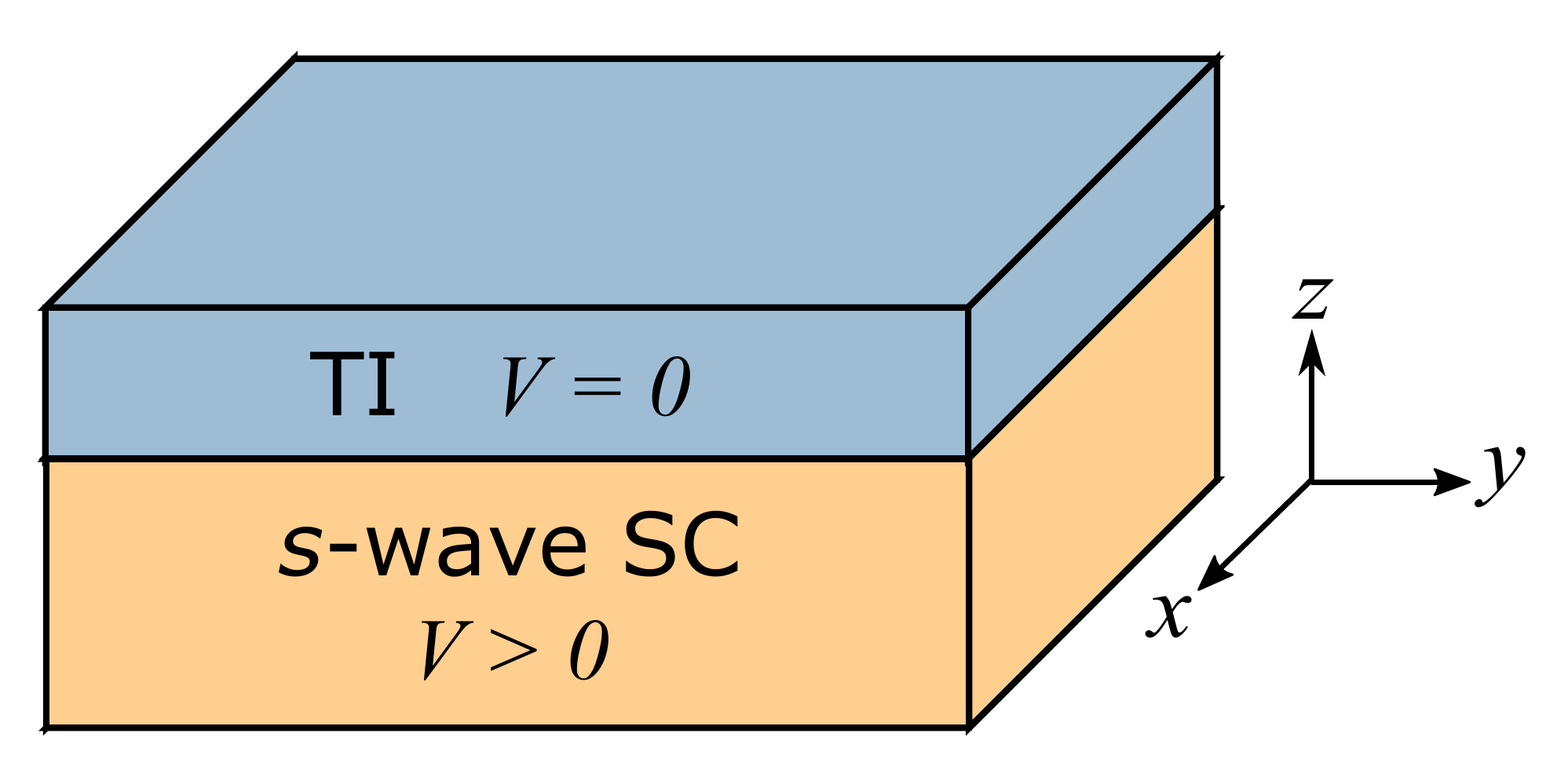}
\caption{ Schematic illustration of the $s$-wave superconductor--3D TI heterostructure that we consider in this work. In our 3D self-consistent simulation, we include a non-zero interaction strength within the $s$-wave superconductor, $V$, but, the interaction strength is initially set to zero in the TI as it contains no natural pairing. The self-consistent procedure to solve for the order parameter then allows for pairing amplitudes from the superconductor to enter the TI and induce superconductivity via the proximity effect. \label{fig:schematic}}
\end{figure}
%=================================================================
%=================================================================

In Fig.~\ref{fig:schematic}, we show a schematic of the $s$-wave superconductor and 3D TI system that we consider in this work. In this heterostructure, we assume the presence of a non-zero attractive interaction, parametrized by the pairing strength $V$, within the superconducting layer and no explicit pairing strength within the TI film. We begin the construction of our model by first considering the non-superconducting Hamiltonian for our heterostructure including the contributions from both the superconductor and the TI. As we wish to accurately model spatially large structures, we write the in-plane directions ($\hat x$ and $\hat y$) within the momentum-space representation while we write the out-of-plane direction ($\hat z$) in real space. In order to maintain notational simplicity, $\vec k$ is used to represent the in-plane momentum $k_x \hat x + k_y \hat y$. The mixed-representation, or that which is composed of both real- and momentum-space components, non-superconducting, nearest-neighbor-hopping Hamiltonian is 
\begin{equation} \label{eq:TB-ham}
\begin{split}
	\mathcal H = \sum_{\vec{k}, z} &\left[ 
		\Psi_{\vec{k},z}^\dag \left(\hat{H}_0(\vec{k},z)\right)\Psi_{\vec{k},z} \right.  \\ 
		&\left. +\left( \Psi_{\vec{k},z}^\dag \hat{H}_z \Psi_{\vec{k},z+\hat z} + \text{H.c.} \right)
		\right], 
\end{split}
\end{equation}
where the annihilation operator $\Psi_{\vec{k},z}=(c_{\vec{k},z,A\uparrow}\; c_{\vec{k},z,A\downarrow}\;c_{\vec{k},z,B\uparrow}\;c_{\vec{k},z,B\downarrow})^T$ is defined by a basis containing two orbitals ($A$, $B$) and two spin degrees of freedom ($\up$, $\dn$). 
The TI Hamiltonian is represented by a minimal bulk model for 3D topological insulator which consists of two spin and two orbital bases\cite{Zhang2010, Chiu2011}, 
\begin{equation} \label{eq:HTI}
\begin{split}
\hat{H}_{\text{3D}}^{\text{latt}}(\vec{k})
=&
M(\vec{k})\Gamma_0 +
\sum_{i=x,y,z} d_i(\vec{k})\Gamma_i + m_z\Gamma_{Z}
\end{split}
\end{equation}
with a momentum space lattice description of
\begin{equation} \label{eq:dM}
\begin{split}
d_i(\vec{k})=&(\hbar v_F/a)\sin(k_i a), \\
M(\vec{k})=&(b/a^2)[\cos(k_x a) +\cos(k_y a)+\cos(k_z a) ] \\
& - 3b/a^2+\mathbb M,
\end{split}
\end{equation}
where $v_F$ is the Fermi velocity, $a$ is a lattice constant, and $b$ is a material parameter used to fit to a specific material band structure. In Eq. (\ref{eq:dM}), $\mathbb M$ is a parameter that controls a topological phase of the system and the system is a trivial (topological) insulator when $\mathbb M/b<0$ ($\mathbb M/b>0$)\cite{Chiu2011}. 
The gamma matrices in Eq. (\ref{eq:HTI}) are defined as $\Gamma_{x,y,z}=\tau_x\otimes\sigma_{x,y,z}$ and $\Gamma_0=\tau_z\otimes I_2$, where $I_N$ are $N \times N$ identity matrices, $\sigma_i$ and $\tau_i$ are the Pauli matrices for spin and orbital degrees of freedom, respectively. The Hamiltonian in Eq. (\ref{eq:HTI}) captures the low-energy characteristics of 3D TRI TIs with the $A$ and $B$ orbitals, for example, that correspond to the linear combination of the $p$ orbitals of Bi$^{3+}$ and Se$^{2-}$ of Bi$_2$Se$_3$, respectively\cite{Zhang2009,Zhang2010}. In addition, we introduce the Zeeman energy, $m_z$, with the corresponding gamma matrix $\Gamma_Z=I\otimes\sigma_z$ in the TI Hamiltonian to capture the magnetization arising from the magnetic dopants in the TI\cite{Zhang2010}  .
To model the $s$-wave superconductor-TI interface of our heterostructure, we represent the conventional superconductor portion of our heterostructure by a simple four-fold degenerate Hamiltonian 
\begin{equation} \label{eq:Hm}
\hat{H}_{\text{m}}^{\text{latt}}(\vec{k})=t_m[3-\cos (k_xa)-\cos (k_ya)-\cos (k_za)] I_4,
\end{equation}
where $t_m$ is the hopping parameter of the system. The metallic Hamiltonian in Eq. (\ref{eq:Hm}) describes a conventional $s$-wave superconductor, for example NbSe$_2$, in the normal phase. Without a loss of generality, we choose $a=b=1$, $t_m=1$, and $\hbar v_F=1$ as our main focus of this work is a qualitative analysis on the order parameter rather than an analysis of an order parameter for a specific material. 
To examine the strong topological insulator regime, we set the mass parameter $\mathbb M=0.35$\cite{Chiu2014}. 
%The resulting model has a bulk band gap of 1.65 eV and a surface state hybridization gap of 221.4 meV. 

In our model, we select the first 10 grid points within our solution space in the $\hat z$-direction to be the metal and the subsequent 4 points are chosen to be 3D TI in order to represent the possible experimental setup in Fig. \ref{fig:schematic}.
%mimic the experimental setup realized in TI thicknesses of 4 quintuple layers.\cite{Xu2014} 
The disparate metal and TI models are incorporated into the full mixed-space Hamiltonian by making the following definitions for the matrices given in Eq. (\ref{eq:TB-ham}):
\begin{widetext}
\begin{align}
	\hat{H}_0(\vec{k}, z) &= \left\{
	\begin{array}{ll}
			(3- \cos k_x - \cos k_y) I_4 - \mu_m I_4 & , z \leq 10 \\
			(\mathbb M -3 + \cos k_x + \cos k_y) \Gamma_0  +  \sin k_x \Gamma_x + \sin k_y \Gamma_y + m_z\Gamma_Z - \mu_{TI} I_4 & , 10 < z \leq 14\\  
	\end{array} \right. \label{eq:H_0}
\\
	\hat{H}_z &= \left\{
	\begin{array}{ll}
		(1/2) I_4 & z < 10 \\
		(1/2) t_c I_4 & z = 10 \\
		(1/2)\Gamma_0 + (i/2) \Gamma_z  & 10 < z \leq 14,
	\end{array} \right. 	\label{eq:H_z}  
\end{align}
\end{widetext}
where $\mu_m$ and $\mu_{TI}$ are the chemical potential of the metallic and TI system, respectively, and $0\leq t_c\leq 1$ is a coupling constant that controls a coupling strength between the metallic and ultrathin TI system.

\begin{figure}[t!]  %============FIGURE=============================
  \centering
   \includegraphics[width=0.5\textwidth]{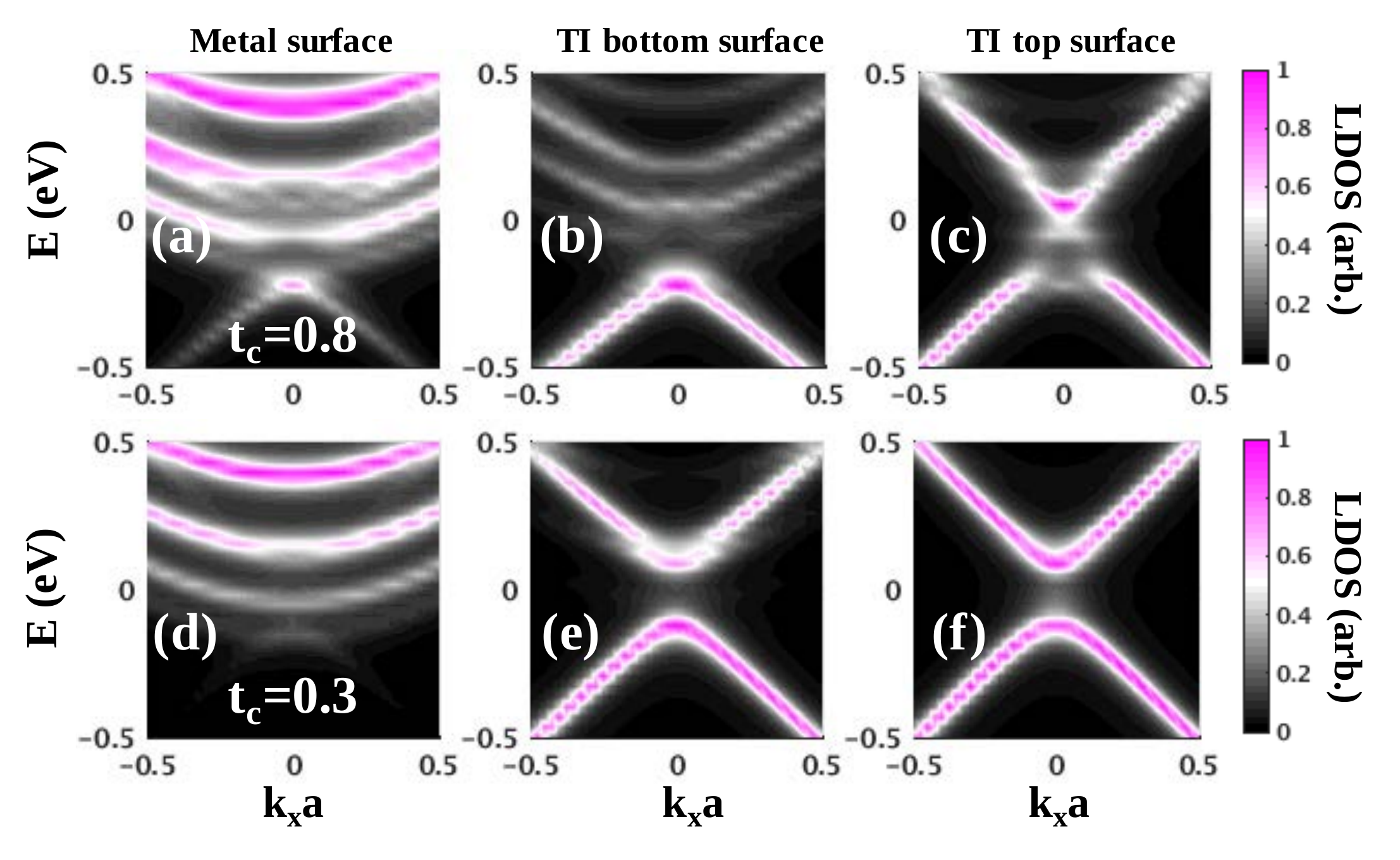}
  \caption{$LDOS(\vec{k},z)$ is plotted at (a) the metallic surface ($z=10$), (b) the bottom surface ($z=11$), and (c) the top surface ($z=14$) for $t_c=0.8$ at $k_y=0$. Similarly, $LDOS(\vec{k},z)$ is plotted at (d) the metallic surface, (e) the bottom surface, and (f) the top surface for $t_c=0.3$. The obtained $LDOS$ is normalized by its maximum magnitude at each layer. }\label{fig:DOS}
\end{figure}   %==========FIGURE=================================
Unlike the isolated TI system, the presence of the metallic states modifies the surface states of the TI when the coupling constant, $t_c$, is introduced. To examine how this coupling modifies the surface states of the TI, we compute the momentum- and real-space resolved local density of states, $LDOS(\vec{k}, z)$, computed by the system Green's function\cite{Datta2005} using Eqs. (\ref{eq:H_0}, \ref{eq:H_z}). Figs. \ref{fig:DOS} (a, b, c) depict the $LDOS(\vec{k}, z)$ at the metallic surface, the bottom layer, and the top layer of the ultrathin TI system, respectively, when a coupling constant is $t_c=0.8$. 
Due to the connecting hopping constant, $t_c\cdot t_m$, whose magnitude is comparable to that of the hopping constant in TI, the surface states of the TI exhibit strong imprint of the electronic states from the metallic states and, as a result, the top surface of the TI shows a reduced hybridization gap.  
In addition, charge accumulation at the interface may occur and cause a structural inversion asymmetry (SIA) potential, which further modifies the surface states band structure of TI\cite{Shen2010}. Such substrate effect and the resultant discrepancy of the band structure from that of the ideal ultrathin TI surface states are observed in, for example, Bi$_2$Se$_3$ system grown on Si(111) substrate\cite{Tjernberg2013} or AlN(0001) substrate\cite{Dimoulas2014}. 
On the other hand, no obvious substrate effect is observed in Bi$_2$Se$_3$ system epitaxially grown on $\alpha-$Al$_2$O$_3$\cite{Kun2011}, GaAs(111)A\cite{Hassan2014}, and conventional $s$-wave superconductor 2H-NbSe$_2$\cite{Xu2014}. In this work, we focus on the moderate to weak coupling regime where the generic band structure of the TI system is minimally affected by the metallic system and we may exclude substrate effects from subsequent analysis. In Figs. \ref{fig:DOS} (d, e, f), we shows the $LDOS(\vec{k}, z)$ at $t_c=0.3$, where we find the TI surface states are close to that of the ideal ultrathin TI system with a well defined hybridization gap. For this reason, we use the coupling constant $t_c=0.3$ for the remainder of this work.

\subsection{Induced $s$-wave order parameter for bulk and surface states of TI and its spatial distribution} \label{sec:3Dantisym}

With the non-superconducting Hamiltonian defined, we may now incorporate the superconductivity into the system. In this work, we include superconductivity at the mean-field level, using an intra-orbital on-site interaction of the form \cite{Hung2013}
\begin{equation}\label{eq:H_int}
\begin{split}
	\mathcal H_\text{int} = -V \!\!\!\!\!\!\ &\sum_{\vec{k}, z, \alpha = A, B} \!\!\!\!\!\!\ \left\{ \Delta_{S,\alpha}^*(z)b_\alpha(\vec{k},z) \right. \\
	&\left. + \Delta_{S,\alpha}(z)b_\alpha^\dagger(\vec{k},z) - \abs{\Delta_{S,\alpha}(z)}^2 \right\}, 
\end{split}
\end{equation}
where $V$ is the on-site attractive interaction strength, and $b_\alpha(\vec{k}, z) = (c_{\vec{k},z,\alpha\up}c_{-\vec{k},z,\alpha\dn} - c_{\vec{k},z,\alpha\dn}c_{-\vec{k},z,\alpha\up})$ is the singlet pair annihilation operator. 
In Eq. (\ref{eq:H_int}), we set $V>0$ in the metal to indicate superconducting pairing whereas $V=0$ in the TI as there is no inherent pairing within the TI.
To examine proximity effect of the TI, we define the $s$-wave order parameter, $\Delta_{S,\alpha}$, as a unitless quantity by separating the interaction strength $V$ from its expression in Eq. (\ref{eq:H_int}). For definitions and numerical calculation procedures for the order parameter in Eq. (\ref{eq:H_int}), see Appendix \ref{app:op2} and Eq. (\ref{eq:DeltaSintra}). 
To incorporate superconductivity into the non-interacting Hamiltonian of Eq. \eqref{eq:TB-ham}, we expand it into a BdG Hamiltonian with corresponding eight-component Nambu spinor $\Phi_{\vec{k}, z} = [\Psi_{\vec{k}, z}, \Psi_{-\vec{k}, z}^\dag]^T $ as\cite{Gennes1966}
\begin{equation} \label{eq:HBdG3D}
\begin{split}
	\mathcal H_\text{BdG} 
		=& \sum_{\vec{k}, z} \Phi_{\vec{k}, z}^\dag 
			\begin{bmatrix}
				\hat{H}_0(\vec{k}, z)   & -V \hat\Delta(\vec{k}, z) \\
				-V \hat\Delta^\dag(\vec{k}, z) &  -\hat{H}_0(-\vec{k}, z)^*
			\end{bmatrix}
			\Phi_{\vec{k}, z} \\  
			& +\Phi_{\vec{k}, z}^\dag 
			\begin{bmatrix}
				\hat{H}_z(z)   &0 \\
				0 &  -\hat{H}_z( z)^*
			\end{bmatrix}
			\Phi_{\vec{k}, z+\hat z}. 
\end{split}
\end{equation}
The intra-orbital $s$-wave interaction term in Eq. \eqref{eq:H_int} is incorporated in the metallic portion of the Hamiltonian through off-diagonal components $\tilde \Delta$, which are written as
\begin{equation}
	\hat\Delta(\vec k, z) = i\sigma_y \otimes \begin{bmatrix}
				\Delta_{S,A}(\vec k, z)   &0 \\
				0 &  \Delta_{S,B}(\vec k, z)
			\end{bmatrix}. \label{eq:tilde_Delta}
\end{equation}
Using the material parameters defined in section \ref{sec:3DTI}, we obtain a hybridization gap of $2\delta\simeq 221$ meV and bulk gap of $\sim1.647$ eV for the 4 layer thick TI Hamiltonian in Eqs. (\ref{eq:H_0}, \ref{eq:H_z}) in normal phase. The chemical potential of the metal is set to $\mu_m=0.2 t_m$ to obtain sufficient density of states near the chemical potential to induce superconductivity within the metallic system. The interaction strength of the metal is set to $V=0.2$ eV to obtain a converged superconducting gap whose magnitude in energy is $\sim 0.24\delta$. From the self-consistent solution of the BdG Hamiltonian given in Eq. (\ref{eq:HBdG3D}), we compute the induced intra-orbital $s$-wave pair correlation function, $F^{\uparrow,\downarrow}_{A,A}(\vec{k},z)=\braket{c_{ \vec{k},z,A\uparrow}c_{ \vec{k},z,A\downarrow}}$, and corresponding $s$-wave order parameter, $\Delta_{S,A}(z)=\sum_{\vec{k}}[F^{\uparrow,\downarrow}_{A,A}(\vec{k},z)-F^{\downarrow,\uparrow}_{A,A}(\vec{k},z)]/2$, within the ultrathin TI system. For detailed numerical calculation procedure, see Appendix \ref{app:op2}. 
Although similar analysis may be done for inter-orbital pairing, we focus on the intra-orbital induced order parameter for this study as an inclusion of the inter-orbital pairing does not differ from the qualitative trends presented here.

\begin{figure}[t!]  %============FIGURE=============================
  \centering
   \includegraphics[width=0.5\textwidth]{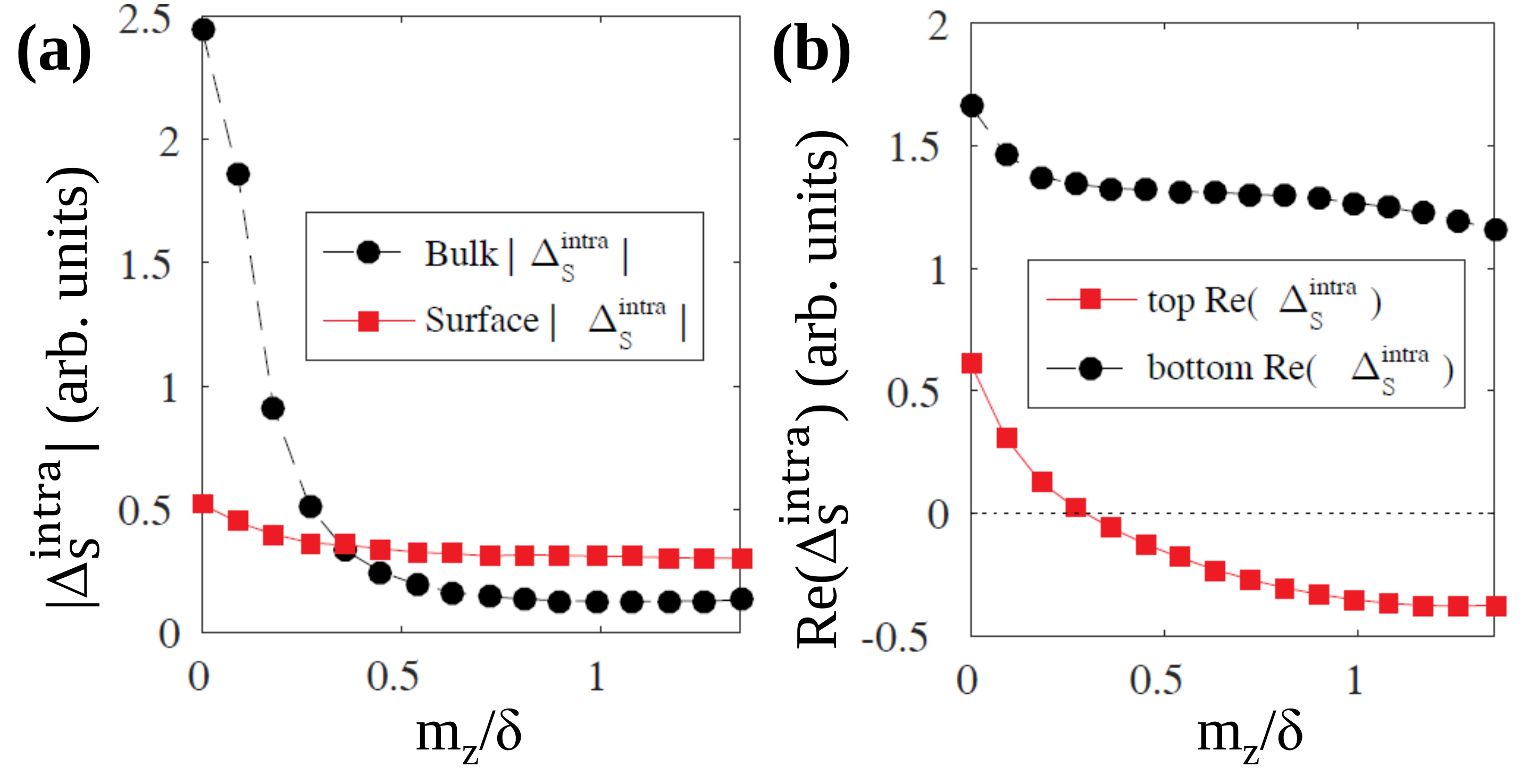}
  \caption{(a) A plot of induced $s$-wave order parameter magnitude as a function of Zeeman splitting energy at the chemical potential $\mu_{TI}=850$ meV, whose location crosses both bulk and surface bands of the TI. Black circles indicate the bulk state contribution and red squares indicate the surface state contribution for the obtained order parameter at the interface of the metal-TI ($z=11$). (b) A plot of the induced $s$-wave order parameter as a function of Zeeman splitting energy at the chemical potential of $\mu_{TI}=\delta$, whose location only crosses the surface bands of the TI. Unlike (a), the induced order parameter is computed solely from the surface states of TI. Black circles indicate the order parameter obtained at bottom layer ($z=11$), whereas the red squares indicate the order parameter at top layer ($z=14$).}\label{fig:OP_BS}
\end{figure}   %==========FIGURE=================================

%%%%%%%%%%%%%%%%%%%%%%%%%%%%%%%%%%%%%%%%%%%%%%%%%%%%%%%%%%%%%%%%%%
%%%%    Delta vs. Mz  (Figure 2)
%%%%%%%%%%%%%%%%%%%%%%%%%%%%%%%%%%%%%%%%%%%%%%%%%%%%%%%%%%%%%%%%%%
\begin{figure*}
	\includegraphics[width=.22\textwidth]{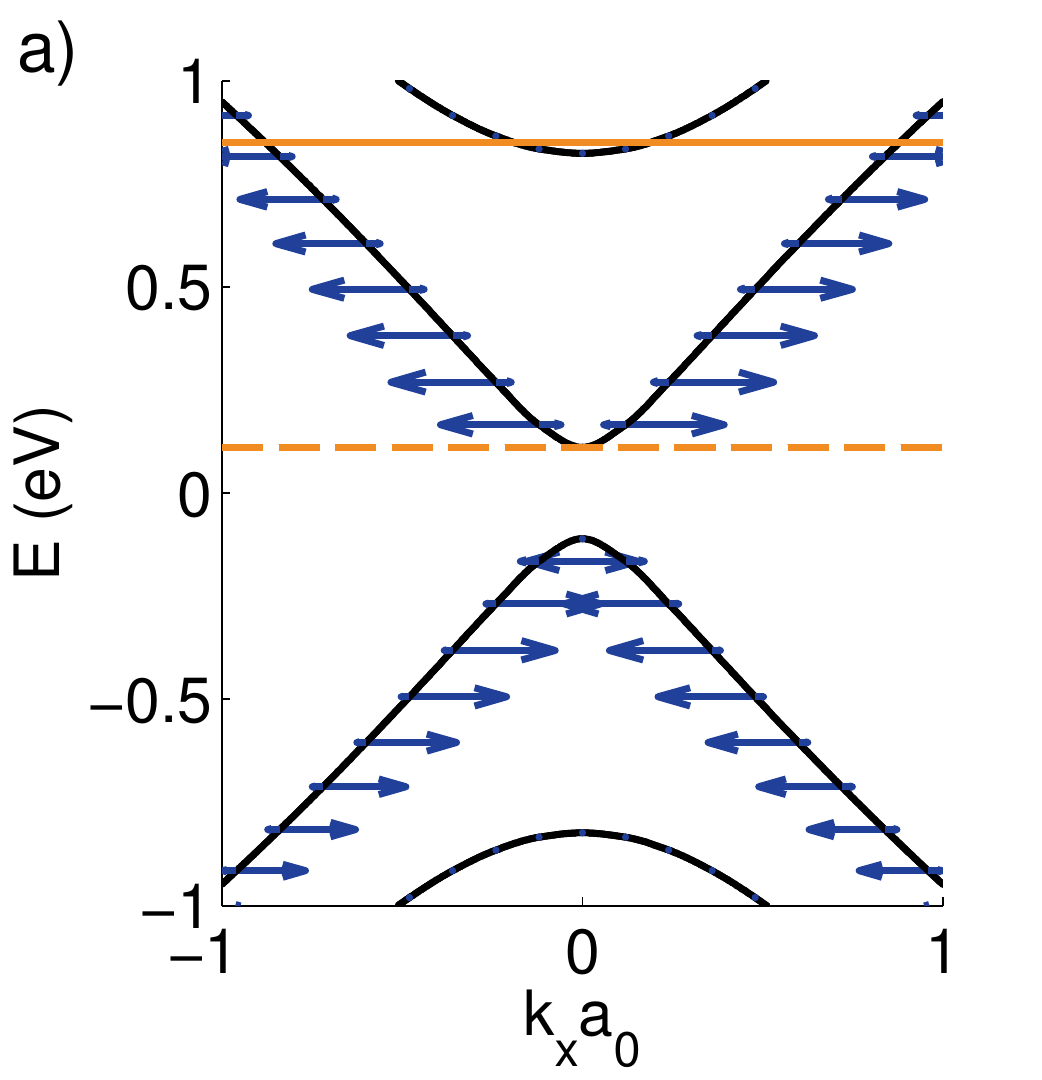}
	\includegraphics[width=.22\textwidth]{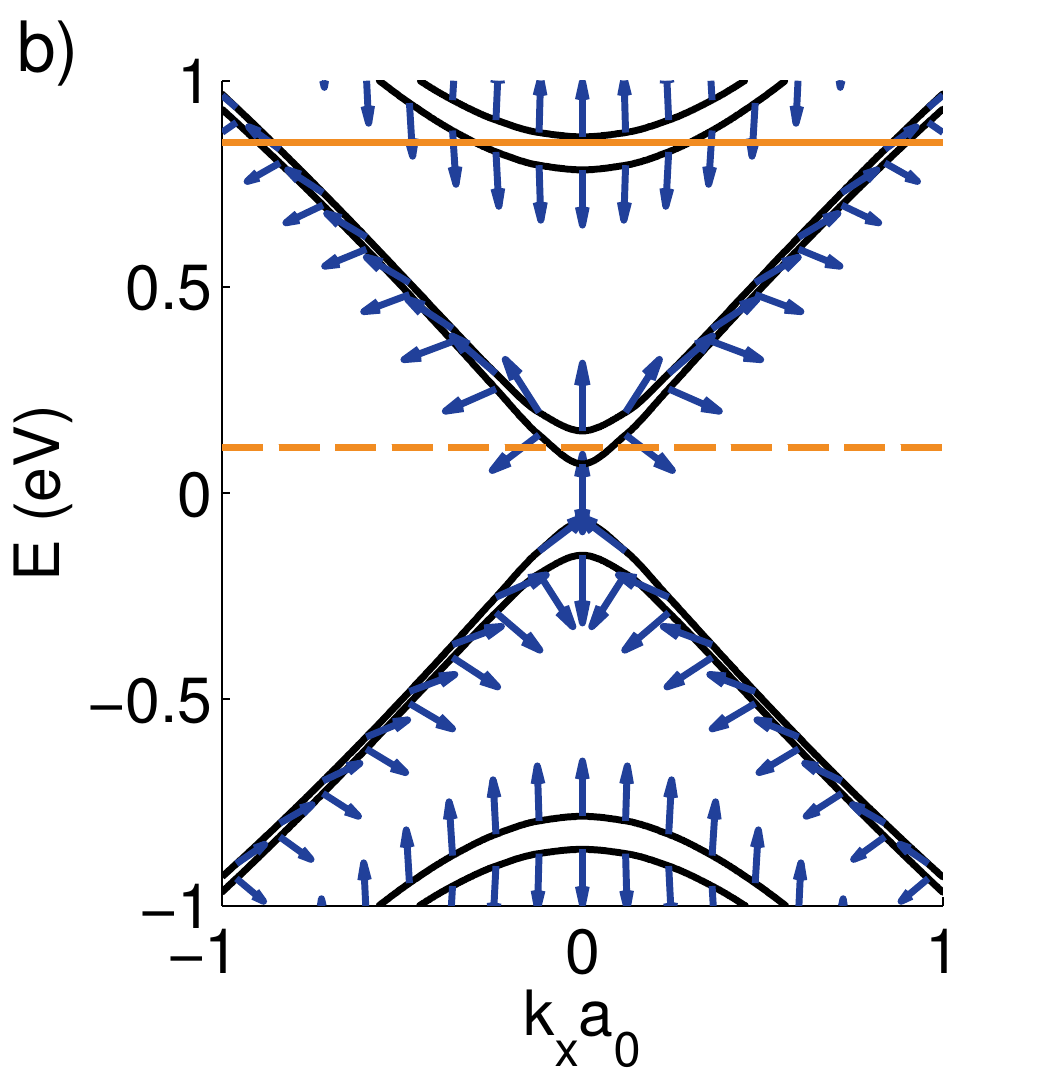}
	\includegraphics[width=.22\textwidth]{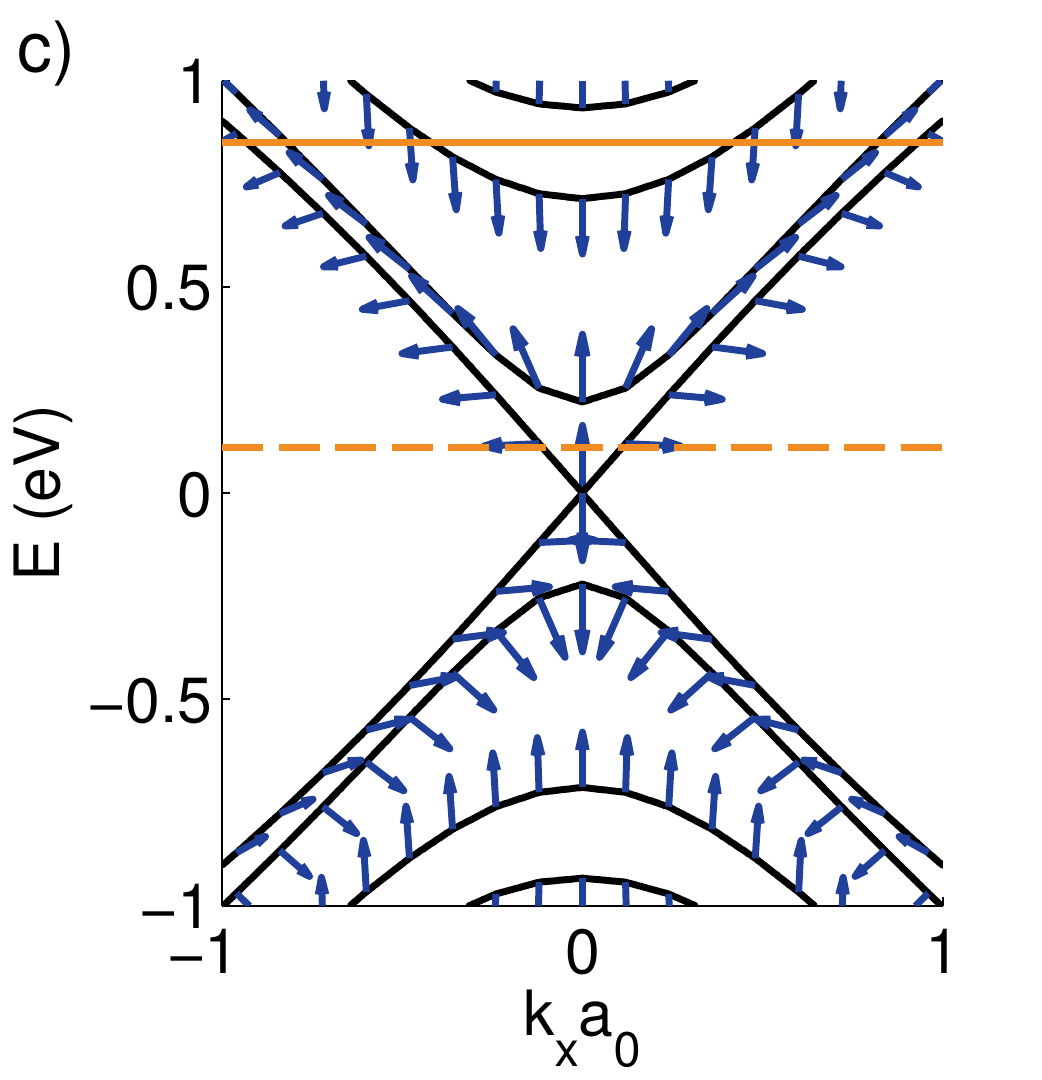}
	\includegraphics[width=.22\textwidth]{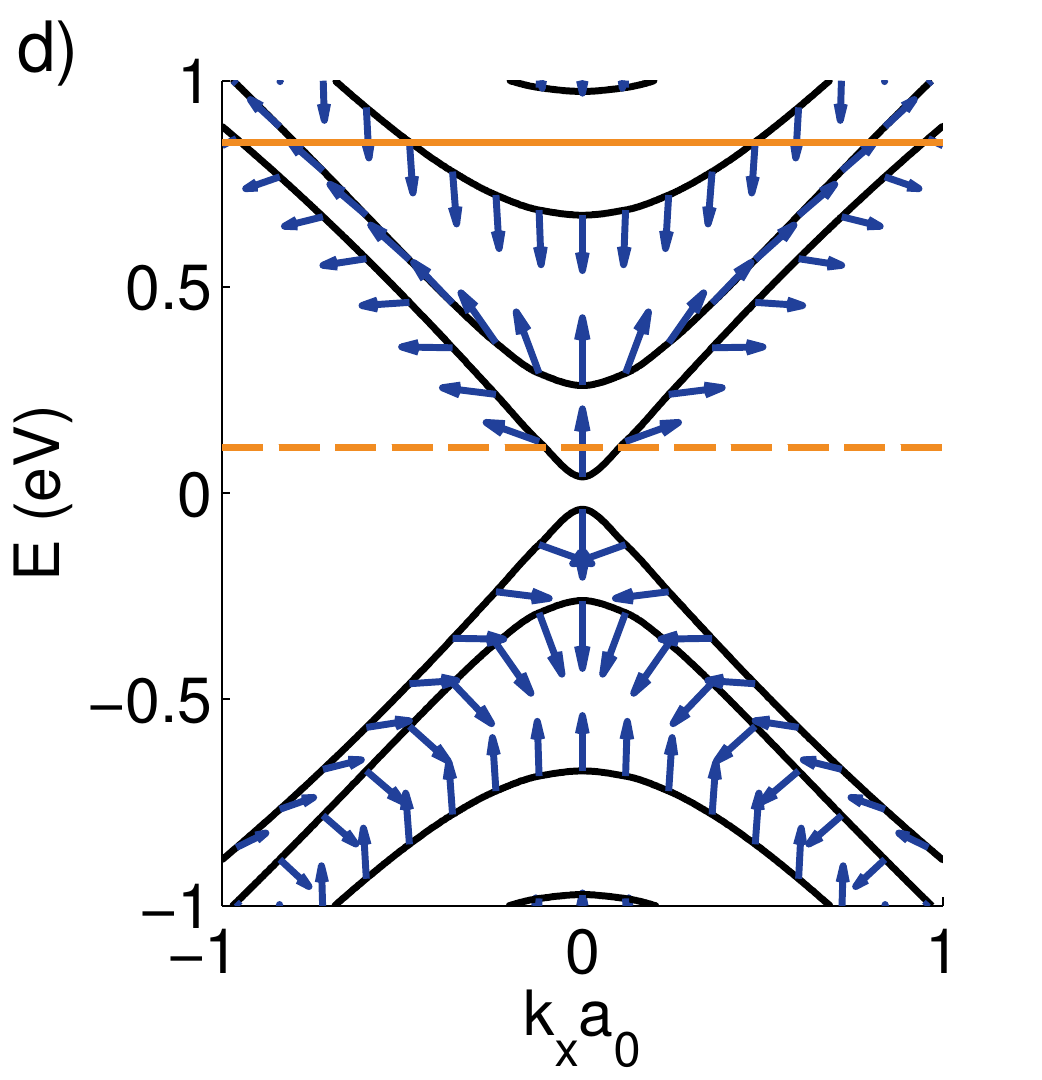}
	\caption{Band structure of the non-superconducting 4 layer thick TI with Zeeman energies of (a) 0 meV, (b) 40 meV, (c) 110 meV, and (d) 150 meV with chemical potentials of 850 meV and 111 meV($=\delta$) marked with solid and dashed horizontal lines, respectively. The spin expectation is marked with blue arrows, where the horizontal component of spin corresponds to $\langle S_y \rangle$ and vertical, $\langle S_z \rangle$.   (a) With no Zeeman field, we see that the surface bands are gapped due to hybridization between the two surfaces. (b) When a small Zeeman term is included, the bulk states split into strictly up and down spins while the surface states' spin cant out of plane. (c) With a Zeeman term of 110 meV, we observe a quantum phase transition into a Chern insulating state\cite{Zyuzin2011}. At this transition, the spin rotates back to an in-plane configuration at small momenta. (d) The Chern insulating state is characterized by the inversion of the band structure, which is apparent in the rotation of spin that is observed within a band when scanning from low to high momenta.
\label{fig:bandstructure}}
\end{figure*}
%=================================================================
%=================================================================

We first examine the induced order parameter for surface and bulk states in the TI and compare its relative magnitude, as well as its evolution as a function of the Zeeman energy. To this end, we locate the chemical potential of the TI at $\mu_{TI}=850$ meV, which crosses both bulk and surface states of the TI. We calculate the magnitude of the intra-orbital $s$-wave order parameter, $|\Delta_S^{\text{intra}}(z)|=|\Delta_{S,A}(z)+\Delta_{S,B}(z)|$, by considering the surface and bulk state contributions separately as a function of the Zeeman splitting energy. The details of how the bulk and surface contributions are separated are discussed in Appendix~\ref{app:3Dkspace}. The induced order parameter is maximized at the interface ($z=11$) as the TI layer is directly coupled to the metallic $s$-wave superconducting system and decays as a function of depth as we move toward the other TI surface ($z=14$). To see a clear trend of the induced order parameter as a function of the Zeeman splitting energy, we examine the magnitude of the induced order parameter at the interface ($z=11$). 
Considering that a typical range of the hybridization gap of $3-5$ quintuple layer Bi$_2$Se$_3$ is $2\delta\sim 138-41$ meV\cite{Xue2010} and magnetic doping induced gap size is $2m_z\sim 87-23$ meV for $\sim 10-0.2$ \% Mn doping in Bi$_2$Se$_3$\cite{Xu2012}, we sweep the Zeeman splitting energy up to $m_z/\delta=1.5$.
As we increase $m_z$, Fig.~\ref{fig:OP_BS}(a) shows a rapid decrease in induced order parameter of the bulk states (black circle), whereas the induced order parameter of the surface states (red square) shows nearly constant magnitude. The seemingly different trend of the bulk and surface states may be explained by understanding the changes in band structure and the spin polarization near the Fermi surface. To explain this, we plot the band structure of a 4 layer thick TI with corrsponding spin texture in Fig.~\ref{fig:bandstructure}, specifically focusing on the states that cross the solid line representing a chemical potential of $\mu_{TI}=850$ meV.  When $m_z = 0$ in Fig.~\ref{fig:bandstructure}(a), we see that the surface states exhibit their spin-locked nature. As the Zeeman field is increased in Figs.~\ref{fig:bandstructure}(b-d), the bulk and surface states show qualitatively different pictures. The bulk states split into strictly up and down spin states, making a spin singlet state energetically unfavorable to form and causing $\Delta_S^{\text{intra}}$ to drop precipitously. The spin of the surface states is markedly different. Because the spin is already locked to momentum at $m_z = 0$, the addition of the Zeeman interaction has the effect of canting the spin out of the \xyplane. However, as there is always a projection of spin that is anti-aligned in \xyplane, $s$-wave pairing is always allowed, explaining why the induced $s$-wave pairing in the surface states decreases much smaller rate than in the bulk. 

We now narrow down our scope to the surface states and examine the induced order parameter at the top and bottom surfaces. Particularly, we change our chemical potential and place it to the bottom of the surface band of the TI. As the introduced Zeeman energy lifts the band degeneracy, this particular position of the chemical potential allows us to focus on a single, non-degenerate surface band and the resultant induced order parameter to elucidate and validate our analysis of the 2D surface model in Section \ref{sec:phenomenological}.
To this end, we locate the chemical potential of the TI at $\mu_{TI}=\delta$ indicated as a dashed line in Fig~\ref{fig:bandstructure}(a). In Fig.~\ref{fig:OP_BS}(b), we find a purely real $s$-wave order parameter from our self-consistent calculation at the top ($z=14$, red square) and bottom ($z=11$, black circle) layer of the ultrathin TI. Both top and bottom layer order parameter exhibit positive sign for zero or small Zeeman splitting energy, implying that symmetric part of the pairing is the dominant factor. As we increase Zeeman energy, we lift the degeneracy to separate the surface bands beyond the superconducting gap and observe a sign change in the induced $s$-wave order parameter at the top layer, implying that the anti-symmetric pairing potential now plays a dominant role over the symmetric pairing potential. Although the overall behavior agrees with the phenomenological model analysis, Fig.~\ref{fig:OP_BS}(b) shows that the spatial distribution of the top and bottom layer induced order parameter is always a mixture of the symmetric and anti-symmetric form. This is due to the fact that the 3D model Hamiltonian takes account the effect of the hopping from the metallic system to the TI. For clarity, we adopt 2D surface model to write an analytic expression of the self-energy term including the hopping from metallic system to bottom surface state of the TI as\cite{Stanescu2010,Sau2010b}
\begin{equation} \label{eq:selfenergy}
\Sigma(\epsilon)\simeq -\frac{\lambda}{\sqrt{\Delta_m^2-\epsilon^2}}(\epsilon\tau'_0+\Delta_m\tau'_x)
\end{equation}
at energy $\epsilon$.
In Eq. (\ref{eq:selfenergy}), the Pauli matrices $\tau'_{i=0,1,2,3}$ act on the Nambu space, $\Delta_m$ is an assumed order parameter of the $s$-wave superconducting system, and $\lambda^{-1}$ is an estimation of the electron life-time at the metal-TI interface which is determined by the hopping constants, $t_c$ and $t_m$, and the chemical potential of the metal. The phenomenological model in Eq. (\ref{eq:BdG}) takes account second term of Eq. (\ref{eq:selfenergy}), however, the first term has been ignored. Assuming that the above self-energy is applied only to the bottom layer of the ultrathin TI due to its direct proximity to the metallic system, we may insert the first term as an on-site potential within the bottom layer Hamiltonian. Specifically, we first define the first term of the self-energy in Eq. (\ref{eq:selfenergy}) as $\Sigma_{1,\epsilon}=-\epsilon\lambda/2\sqrt{\Delta_m^2-\epsilon^2}$. Utilizing the symbols defined in Eq. (\ref{eq:Hsurf}) and below, we incorporate the first term of the self-energy in Eq. (\ref{eq:selfenergy}) with the 2D surface Hamiltonian in Eq. (\ref{eq:Hsurf}) as an on-site potential at bottom layer, $\hat{H}_{BdG,\Sigma_{1,\epsilon}}=\Sigma_{1,\epsilon}\tau'_0\otimes(I_2\otimes I_2-\tau_z\otimes I_2)=\tau'_0\otimes \hat{H}_{\Sigma_{1,\epsilon}}$. Following the discussion in Eq. (\ref{eq:Hsurf2}), we apply the relevant rotational matrices to obtain a surface Hamiltonian
\begin{equation}
\hat{H}'_{surf}(k,\epsilon)=U_2 U_1 [\hat{H}_{surf}(k_x,k_y)+\hat{H}_{\Sigma_{1, \epsilon}}]U_2^\dagger U_1^\dagger,
\end{equation}
whose matrix form is
\begin{equation}  \label{eq:Hsurfomega}
\hat{H}'_{\text{surf}}(k,\epsilon)=
\begin{pmatrix}
m_2-\mu' & 0 & -\Sigma_{1,\epsilon} & -k \\
0 & m_1-\mu' & -k & -\Sigma_{1,\epsilon} \\
-\Sigma_{1,\epsilon}& -k & -m_1-\mu' & 0 \\
-k & -\Sigma_{1,\epsilon} & 0 & -m_2-\mu' \\
\end{pmatrix},
\end{equation}
where $\mu'=\mu+\Sigma_{1,\epsilon}$. Note that the off-diagonal term, $\Sigma_{1,\epsilon}$, in Eq. (\ref{eq:Hsurfomega}) couples the sectors 1 and 2 which were previously decoupled in Eq. (\ref{eq:Hsurf22}). Considering the correlation function near the chemical potential, or $\epsilon\sim0$, the additional coupling term $\Sigma_{1,\epsilon}$ has a small but non-zero value. Therefore, the presence of the coupling between sector 1 and sector 2 leads the co-existence of the symmetric and anti-symmetric components of the induced order parameter, as we no longer can decouple sector 1 and sector 2 completely. As a result, Fig.~\ref{fig:OP_BS}(b) exhibits the spatial distribution of the induced order parameter as a mixture of the symmetric and anti-symmetric form and exhibits a smooth transition of the pairing potential from symmetric to anti-symmetric form. Nevertheless, we clearly observe a transition of the induced $s$-wave order parameter from a symmetric dominant to an anti-symmetric dominant spatial distribution and thereby confirm that qualitative behavior of the induced order parameter is sufficiently captured at the phenomenological level.

\section{Summary and Conclusion} \label{sec:conclusion} 
In summary, we studied magnetically-doped ultrathin TI system that is proximity coupled to a conventional $s$-wave superconductor in order to identify possible topological phases. 
We find that the system is described in two individual sectors comprised of the hybridized basis with top and bottom surfaces of TI. Using a simplified picture for proximity induced order parameter, we identify that a symmetric and anti-symmetric spatial configuration of the induced order parameter pair electrons in the individual sectors differently. Our subsequent analysis of the total energy of this system reveals that the anti-symmetric spatial distribution of the induced order parameter is dominant in the presence of a Zeeman energy larger than the magnitude of the induced order parameter. As the choice of an anti-symmetric order parameter greatly simplifies the analysis, we perform analytic analysis on the quasi-particle spectrum of the BdG Hamiltonian. We then find the gap closing points at $\vec{k}=0$ and identify that the gap closing points are modified by the hybridization gap, Zeeman energy, and chemical potential of the ultra-thin TI. Taking account both symmetric and anti-symmetric pairing potential, we draw a generic phase diagram to identify a relevant region in parameter space for topological superconductivity. 
To enhance our understanding on proximity coupled ultrathin TI system, we study a more realistic model with a four band 3D TI Hamiltonian directly coupled with metallic superconducting system. In this system, the induced order parameter is determined self-consistently. 
We confirm that our phenomenological model captures the correct behavior of the induced order parameter as the self-consistently determined order parameter shows a clear transition from symmetric to anti-symmetric dominant spatial distribution form. Furthermore, the results show that the surface state induced $s$-wave order parameter survives even at a relatively high magnitude of Zeeman energy, whereas the bulk state induced $s$-wave order parameter exhibits a rapid decay for the increased Zeeman energy. 
We believe that this work not only explains why topological superconductivity yet to be observed in this system, but also provides useful experimental insight into the proper choice of parameters where it may be found.

\begin{acknowledgments} 
T.M.P., M.J.P., and M.J.G. acknowledge financial support from the National Science Foundation (NSF) under grant CAREER ECCS-1351871.
\end{acknowledgments}

%==============================================
\pagebreak
\clearpage
\onecolumngrid

\appendix
\onecolumngrid
\setcounter{equation}{0}
\renewcommand{\theequation}{A\arabic{equation}}
\subsection{The correlation function of the s-wave order parameter for 2D surface model} \label{app:op}

In this section, we describe how the induced order parameter in 2D surface state model of the TI is calculated.
Within the phenomonological model for the induced order parameter, we consider the $8\times 8$ 2D BdG Hamiltonian defined in Eq. (\ref{eq:BdG}) in the main text. The particle operator is defined as $\Phi_\vec{k}=[\Psi_{\vec{k}}, \Psi_{-\vec{k}}^*]^{T}$ where $\Psi=[c_{t,\uparrow}(\vec{k}),c_{t,\downarrow}(\vec{k}),c_{b,\uparrow}(\vec{k}),c_{b,\downarrow}(\vec{k})]$. We diagonalize the Hamiltonian in Eq. (\ref{eq:BdG}) and obtain eigenfunctions, $U$, and eigenvalues, $D$, which satisfies $\hat{H}_{\text{BdG}}U=UD$. Then we define the quasi-particle operator $\Gamma_{\vec{k}}=U^\dagger \Phi_{\vec{k}}$ from 
$\Phi_\vec{k}^\dagger \hat{H}_{\text{surf}}\Phi_\vec{k}
=\Phi_\vec{k}^\dagger U (U^\dagger \hat{H}_{\text{surf}} U) U^\dagger\Phi_\vec{k}
=\Gamma_{\vec{k}}^\dagger D \Gamma_{\vec{k}}$, where $\Gamma_{\vec{k}}=[\gamma_{1,\vec{k}}, \dots, \gamma_{8,\vec{k}}]^T$ with the quasi-particle spectrum index ranging from $1,2,\cdots,8$. The particle operator may be written as Bogoliubov quasi-particle operators, namely, $\Phi_{\vec{k}}=U\Gamma_{\vec{k}}$, where each element is expressed as $[\Phi_{\vec{k}}]_i=\sum_j [U]_{ij}[\Gamma_{\vec{k}}]_{j}$.
For example, the electron annihilation operator of the top surface is
\begin{equation}
c_{t,\uparrow}(\vec{k})=[\Phi_{\vec{k}}]_1=\sum_i[U]_{1i}[\Gamma]_{i}
=\sum_iu_{1i}\gamma_{i,\vec{k}},
\end{equation}
where we define $u_{ij}=[U]_{ij}$.
As a result, we may write the correlation function of the top surface as
\begin{equation} \label{eq:UfU}
\begin{split}
&F_{t}^{\uparrow\downarrow}(\vec{k})=\expect{c_{t,\uparrow}(\vec{k}) c_{t,\downarrow}(-\vec{k})}
=\sum_{i,j} u_{1i}u_{6j}^\dagger\expect{\gamma^\dagger_{j,\vec{k}}\gamma_{i,\vec{k}} } \\
&=\sum_i u_{1i}u_{6i}^\dagger\expect{\gamma^\dagger_{i,\vec{k}}\gamma_{i,\vec{k}} }
= \sum_i u_{1i}u_{6i}^\dagger f(E_{i,\vec{k}})  \\
&=[U\cdot f(D) \cdot U^\dagger]_{16},
\end{split}
\end{equation}
where $f(E)$ is Fermi-Dirac distribution at energy $E$, and we utilize the quasi-particle operator properties: $\gamma_i^\dagger\gamma_j+\gamma_j^\dagger\gamma_i=\delta_{ij}$ and $\expect{\gamma_i^\dagger\gamma_i}=f(E_i)$.
Finally, the correlation function of the s-wave order parameter at the top surface is defined as
\begin{equation}
\begin{split}
F_{S,t}(\vec{k})=&F_{t}^{\uparrow\downarrow}(\vec{k}) - F_{t}^{\downarrow\uparrow}(\vec{k}) \\
=&[U\cdot f(D) \cdot U^\dagger]_{16} - [U\cdot f(D) \cdot U^\dagger]_{25}. \\
\end{split}
\end{equation}
The correlation function of the bottom surface is determined similarily.

\subsection{Numerical evaluation of the Chern number in a 2D square lattice} \label{app:Chern}
In this section, we describe how the Chern number is calculated in a 2D square lattice.
We begin with the generic Hamiltonian having its eigenfucntion of the $n$th band, $\ket{n(\vec{k}_l)}$, where the momentum is defined in a square lattice Brillouin zone as
\begin{equation}
\vec{k}=(k_{j_x}, k_{j_y}), \; k_{j_{x/y}}=\frac{2\pi j_{x/y}}{N_{x/y}},
\end{equation}
with the momentum index $j_{x/y}=0,\cdots,N_{x/y}-1$ and $l=1, \cdots, N_xN_y$. Note that the eigenfunction is periodic in momentum space and satisfies $\ket{n(\vec{k}_l)}=\ket{n(\vec{k}_l+N_i\hat{e}_i)}$, where $\hat{e}_{i}$ is a vector in the $i=(x,y)$ direction.
To compute the field strength through a lattice plaquette, we first define a link variable\cite{Hatsugai2005} that corresponds to the Berry connection in reciprocal space in the continuum as
\begin{equation} \label{eq:U1link}
U_i(\vec{k}_l)=\bra{n(\vec{k}_l)} n(\vec{k}_l+\hat{e}_i)\rangle/\mathcal{N}_i(\vec{k}_l),
\end{equation}
assuming that $\mathcal{N}_i(\vec{k}_l)=|\bra{n(\vec{k}_l)} n(\vec{k}_l+\hat{e}_i)\rangle|$ is well-defined. Then the field strength is evaluated as\cite{Hatsugai2005}
\begin{equation} \label{eq:Fxy}
\tilde{C}_{xy}(\vec{k}_l)=\ln 
U_x(\vec{k}_l)U_y(\vec{k}_l+\hat{x})
U_x(\vec{k}_l+\hat{y})^{-1}U_y(\vec{k}_l)^{-1},
\end{equation}
which satisfies $-\pi<\tilde{C}_{xy}(\vec{k}_l)/i \leq \pi$ due to the logarithm applied to the link variables. By summing over all the plaquettes, we obtain the Chern number of $n$th band as
\begin{equation}
\tilde{c}_n=\frac{1}{2\pi i}\sum_l \tilde{C}_{xy}(\vec{k}_l).
\end{equation}

\subsection{3D TI induced order parameter calculation} \label{app:op2}
In this section, we describe how the induced order parameter of the 3D TI model is calculated.
In order to self-consistently solve the system of equations, we diagonalize the BdG Hamiltonian to determine the spectrum of the proximity-coupled system. To simplify this procedure, the Bogoliubov transform is taken on Eq. \eqref{eq:HBdG3D} to form the eigenvalue equation
\begin{align}
	\hat{H}_\text{BdG}(\vec k, z) &\begin{bmatrix}
					u_{\vec k, z, n}   & -v_{\vec k, z, n}^* \\
					v_{\vec k, z, n} &  u_{\vec k, z, n}^*
				\end{bmatrix}  = \begin{bmatrix}
					E_n   & 0 \\
					0 &  -E_n
				\end{bmatrix}
				\begin{bmatrix}
					u_{\vec k, z, n}   & -v_{\vec k, z, n}^* \\
					v_{\vec k, z, n} &  u_{\vec k, z, n}^*
				\end{bmatrix}, \label{eq:eig_eq}
\end{align}
where we suppress spin and orbital index for simplicity and the energy eigenvalues $E_n$ correspond to the eigenvector $\psi_{n,\vec k, z} = \{ u_{\vec k, z, n},v_{\vec k, z, n} \}^T$. Then a pair correlation function is defined and may be computed numerically from this diagonalization procedure as following:
\begin{equation} \label{eq:Delta}
	F_{\alpha,\alpha'}^{s,s'} (\vec k, z) =\braket{c_{\vec k,z,\alpha,\sigma}c_{-\vec k,z,\alpha',\sigma'}}= \sum_n u_{\vec k, z, n}^{\alpha,s}(v_{\vec k, z, n}^{\alpha',s'})^* [1-2f(E_n)], 
\end{equation}
where $\alpha,\alpha'$ are orbital index, $s,s'$ are spin index, $f(E)$ is Fermi-Dirac distribution. We note that the selection of a temperature in our system is of little relevance to the simulations so long as it is smaller than the size of the superconducting gap. Threfore, we set the temperature to $T = 0$ K.

In this work, we consider Hubbard type on-site interaction for spin-singlet pairing. To consider relevant electron-electron pairing, we consider an on-site interaction for spin-singlet pairing channel:
\begin{equation} \label{eq:intST}
\hat{H}_{\text{int}}=-\sum_{\vec{r},\alpha,\alpha',s,\bar{s}}V_{\vec{r}} n_{\vec{r},\alpha,s}n_{\vec{r},\alpha',\bar{s}},
\end{equation}
where $V_{\vec{r}}>0$ represents interaction strength, $n_{\vec{r},\alpha,s}=c^\dagger_{\vec{r},\alpha,s}c_{\vec{r},\alpha,s}$ is the density operator to describe the occupation number on $\alpha$ orbital with spin state $s(\neq \bar{s})$ at site $\vec{r}$.
Assuming a constant interaction strength, $V_\vec{r}=V>0$, the mean field decomposition of the spin-singlet pairing Hamiltonian in Eq. (\ref{eq:intST}) results in
\begin{equation} \label{eq:HSreal}
\begin{split}
\hat{H}_{\text{int}}^{\text{MF}}=&
-V\sum_{\vec{r},\alpha,\alpha'} \left\{
F_{\alpha,\alpha'}^{\uparrow,\downarrow}(\vec{r},\vec{r})
c^\dagger_{\vec{r},\alpha',\downarrow}c^\dagger_{\vec{r},\alpha,\uparrow}  +
F_{\alpha,\alpha'}^{\downarrow,\uparrow}(\vec{r},\vec{r})
c^\dagger_{\vec{r},\alpha',\uparrow}c^\dagger_{\vec{r},\alpha,\downarrow}
+ h.c. 
- F_{\alpha,\alpha'}^{\uparrow,\downarrow}(F_{\alpha,\alpha'}^{\uparrow,\downarrow})^\dagger
- F_{\alpha',\alpha}^{\downarrow,\uparrow}(F_{\alpha',\alpha}^{\downarrow,\uparrow})^\dagger
\right\} \\
=&
-V\sum_{\vec{r},\alpha,\alpha'} \left\{
\Delta_{S,\alpha\alpha'}(\vec{r})c^\dagger_{\vec{r},\alpha',\downarrow}c^\dagger_{\vec{r},\alpha,\uparrow}  -
\Delta_{S,\alpha\alpha'}(\vec{r})c^\dagger_{\vec{r},\alpha,\uparrow}c^\dagger_{\vec{r},\alpha',\downarrow}
+ h.c. 
- F_{\alpha,\alpha'}^{\uparrow,\downarrow}(F_{\alpha,\alpha'}^{\uparrow,\downarrow})^\dagger
- F_{\alpha',\alpha}^{\downarrow,\uparrow}(F_{\alpha',\alpha}^{\downarrow,\uparrow})^\dagger
\right\}, \\
\end{split}
\end{equation}
where we define the pair correlation function in real space as
\begin{equation}\label{eq:Freal}
F_{\alpha,\alpha'}^{s,s'} (\vec{r},\vec{r}')=\braket{c_{\vec{r},\alpha,\sigma}c_{\vec{r}',\alpha',\sigma'}},
\end{equation} 
and a spin-singlet order parameter in real space, 
\begin{equation}
\Delta_{S,\alpha\alpha'}(\vec{r})=\frac{1}{2}[F_{\alpha,\alpha'}^{\uparrow,\downarrow} (\vec{r},\vec{r}) - F_{\alpha',\alpha}^{\downarrow,\uparrow} (\vec{r},\vec{r})].
\end{equation}

We perform a Fourier transform of Eq. (\ref{eq:HSreal}) to momentum space in $\hat{x}$ and $\hat{y}$ direction. As a result, we obtain 
\begin{equation}
\begin{split}
\hat{H}_{\text{int}}^{\text{MF}}=& 
-V\sum_{\vec{k},z,\alpha}  \left\{
\Delta_{S,\alpha}(z) c^\dagger_{-\vec k,z,\alpha,\downarrow}c^\dagger_{\vec k,z,\alpha,\uparrow} +
\Delta_{S,\alpha,\bar\alpha}(z) c^\dagger_{-\vec k,z,\bar\alpha,\downarrow}c^\dagger_{\vec k,z,\alpha,\uparrow} + h.c. \right\},
\end{split}
\end{equation}
where we drop the constant term from Eq. (\ref{eq:HSreal}). Then the intra-orbital singlet order parameter is defined as
\begin{equation} \label{eq:DeltaSintra}
\Delta_{S,\alpha}(z)=\frac{1}{2}\sum_{\vec{k}} \;
[F_{\alpha,\alpha}^{\uparrow,\downarrow}(\vec k, z)  - F_{\alpha,\alpha}^{\downarrow,\uparrow}(\vec k, z) ],
\end{equation}
and the inter-orbital singlet order parameter is defined as
\begin{equation} \label{eq:DeltaSinter}
\Delta_{S,\alpha\bar{\alpha}}(z)=\frac{1}{4}\sum_{\vec{k}} \;
[F_{\alpha,\bar{\alpha}}^{\uparrow,\downarrow}(\vec k, z)  
- F_{\bar{\alpha},\alpha}^{\downarrow,\uparrow}(\vec k, z)
+F_{\bar{\alpha},\alpha}^{\uparrow,\downarrow}(\vec k, z)  
- F_{\alpha,\bar{\alpha}}^{\downarrow,\uparrow}(\vec k, z) ], 
\end{equation}
where $\bar\alpha\neq\alpha$, and each of the Fourier transformed pair correlation functions are computed numerically using Eq. (\ref{eq:Delta}). Note that Eqs. (\ref{eq:DeltaSintra}) and (\ref{eq:DeltaSinter}) are even under momentum exchange and odd under spin exchange.

%%%%%%%%%%%%%%%%%%%%%%%%%%%%%%%%%%%%%%%%%%%%%%%%%%%%%%%%%%%%%%%%%%
%%%%    Bands + triplet breakdown  Mz = 0D0 (Figure 3)
%%%%%%%%%%%%%%%%%%%%%%%%%%%%%%%%%%%%%%%%%%%%%%%%%%%%%%%%%%%%%%%%%%
\begin{figure}[t]
\includegraphics[width=0.7\textwidth]{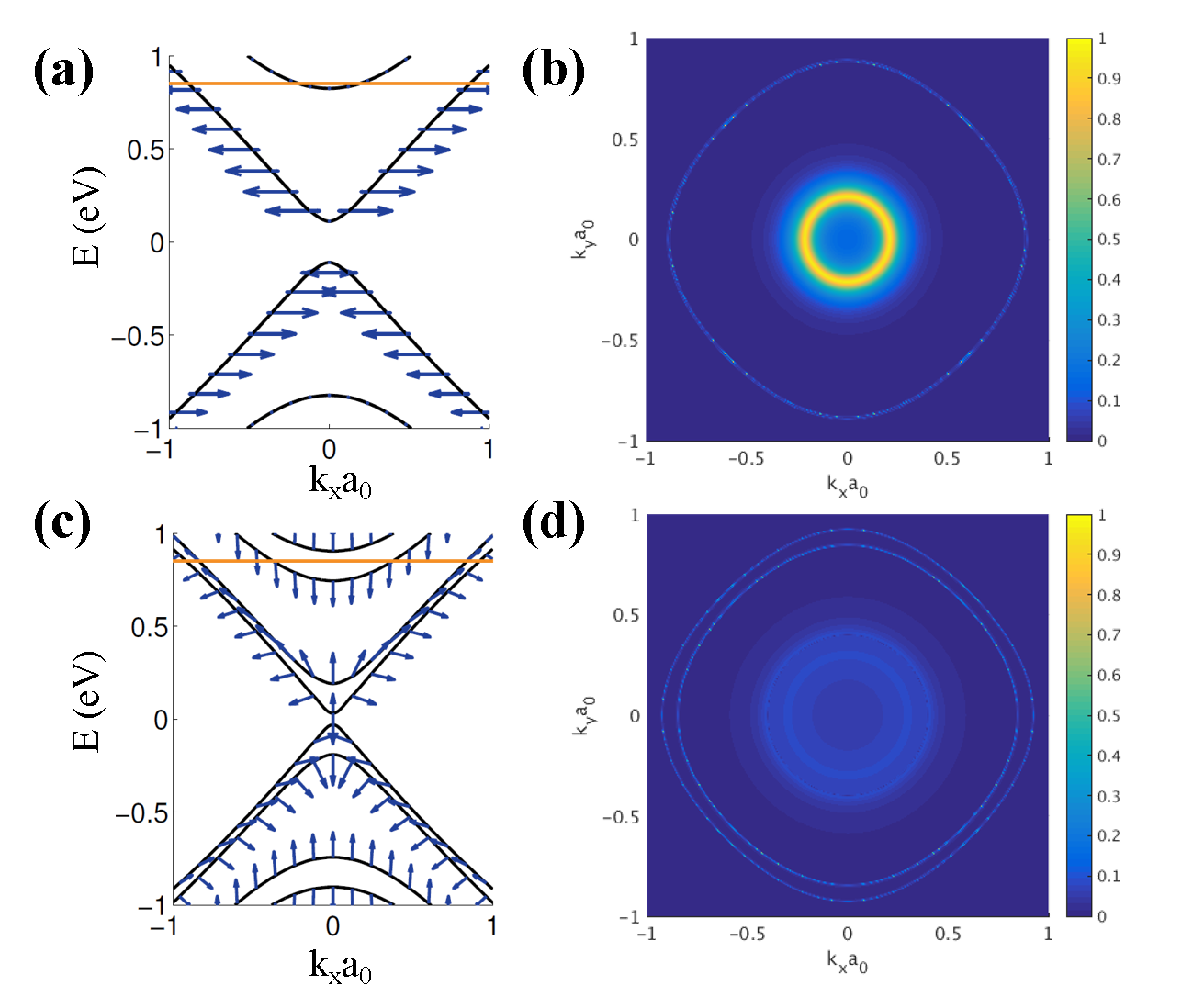}
\caption{(a) Band structure of the top surface ($z=14$) of a non-superconducting 4-unit-cell TI at $m_z =  0$ and $k_y = 0$ with the chemical potential of {850} meV as indicated by the solid line. Spin is indicated with blue arrows, and for visual clarity, the horizontal axis for the arrows corresponds to the $\hat y$-component of spin, while the vertical axis is for $\hat z$-component of spin. (b) The magnitude of the momentum resolved $s$-wave order parameter magnitude $|\Delta_S^{\text{intra}}(\vec{k},z=14)|$ defined in Eq. (\ref{eq:kresolveddelta}). The result displays a clear bulk contribution in the Brillouin zone center and a surface state contribution in the outer ring. (c) Band structure of the top surface of a non-superconducting 4-unit-cell TI at $m_z$=  80 meV and $k_y = 0$  with the chemical potential of {850} meV as indicated by the solid line. We observe that the spin-degenerate bulk states split due to the presence of the magnetization. In the surface states, the Zeeman energy instead cants the spin out of the plane into the $\pm\hat{z}$-direction.  (d) The bulk contribution to the $s$-wave component of the order parameter is strongly suppressed due to the presence of $\hat{z}$-directed magnetism while the surface contribution is resilient as the canted spins still have a significant in-plane projection that allows for persistent $s$-wave pairing. }\label{fig:Bands_OP}
\end{figure}
%%%%%%%%%%%%%%%%%%%%%%%%%%%%%%%%%%%%%%%%%%%%%%%%%%%%%%%%%%%%%%%%%%
%%%%%%%%%%%%%%%%%%%%%%%%%%%%%%%%%%%%%%%%%%%%%%%%%%%%%%%%%%%%%%%%%%  

%======================================================================
%======================================================================
%======================================================================
\subsection{Momentum-space mapping of order parameter in 3D heterostructure}\label{app:3Dkspace}

A wealth of information can be gathered from analyzing the momentum resolution we have in our self-consistent 3D simulation by defining the momentum-resolved intra-orbital $s$-wave order parameter
\begin{equation} \label{eq:kresolveddelta}
\Delta_{S}^{\text{intra}}(\vec{k},z)=
\frac{1}{2}[
F_{A,A}^{\uparrow,\downarrow}(\vec k, z)  
- F_{A,A}^{\downarrow,\uparrow}(\vec k, z)
+F_{B,B}^{\uparrow,\downarrow}(\vec k, z)  
- F_{B,B}^{\downarrow,\uparrow}(\vec k, z)].
\end{equation}
We first consider these features when time-reversal symmetry is preserved with $m_z = 0$ in Figs.~\ref{fig:Bands_OP} (a-b). In Fig.~\ref{fig:Bands_OP}(a), we plot the band structure of the top-most layer of a 4-unit-cell TI without proximity coupling or superconductivity. The $\hat y$ and $\hat z$ directional spin texture are illustrated as the horizontal and vertical vector components, respectively, of the superimposed arrows, whose length is proportional to the relative magnitude of the individual components. In the surface band structure, we can clearly observe both the hybridization gap and the spin-momentum locking of the topologically non-trivial surface bands. The bulk states are spin-degenerate, therefore, we see no net spin texture in bulk bands. The chemical potential indicated by the solid, horizontal line shows that both bulk and surface states are available for pairing at the Fermi surface. By noting the intersection of the chemical potential with the energy bands of the non-superconducting TI, we can identify the bulk and surface components of the $s$-wave order parameter of the proximity-coupled heterostructure in Fig.~\ref{fig:Bands_OP}(b). The large contribution in the inner ring corresponds to the bulk component of $s$-wave pairing. Similarly, the large outer ring in $\Delta_S^{\text{intra}}(\vec{k},z=14)$ corresponds to the wave vectors associated with the surface states and their associated $s$-wave contribution. This knowledge allows us to unambiguously identify the surface and bulk contributions analyzed in the main text. Specifically, we separate surface states contribution from bulk states contribution by computing Eq. (\ref{eq:DeltaSintra}) using the correlation function obtained at momentum space points that satisfy $|\vec{k}|>|\vec{k_c}|$. In this work, we set the momentum cut-off as $|\vec{k_c}a|=0.6$ in the determination of the $s$-wave order parameters for bulk and surface states.

With the physics of the order parameter understood when time-reversal symmetry is preserved, Figs.~\ref{fig:Bands_OP} (c-d) now show the effects of Zeeman energy on the resulting superconducting order parameter when $m_z = {80}$ meV, which explicitly breaks  time-reversal symmetry. The effect of Zeeman energy is evident in the non-superconducting TI surface band structure, shown in Fig.~\ref{fig:Bands_OP}(c). When $m_z > 0$, we observe that the spin-degenerate bulk bands split with spin oriented completely in the $\hat z$ direction with the up-spin states are pushed to higher energies and the down-spin states to lower. In the surface bands, we see a qualitatively different picture. Because of the initial in-plane spin configuration of the hybridized surface bands, $m_z> 0$ has the effect of canting the spin texture towards the $\pm\hat z$, or, out-of-plane direction. This canted spin texture is characteristic of the Zeeman splitting of surface bands that has been observed experimentally in the presence of magnetically ordered impurities in TI samples\cite{Xu2012}. Due to the hybridization of the surface state wave functions in the system, the surface bands also split in energy with the energy change dictated by the sign of $z$-directed cant: up-canted spins rise in energy and down-canted spins lower. The presence of magnetism has immediate consequences to the different components of the superconducting order parameter. In Fig.~\ref{fig:Bands_OP}(d), we observe that because the up- and down-spin states in the bulk are no longer present at opposing momenta, the bulk contribution to $s$-wave pairing is strongly suppressed. The surfaces states however, still have strong $s$-wave pairing despite the Zeeman splitting from the magnetism. Due to the fact that the spins of the surface states are only canted, a projection of spin is still anti-parallel across the Brillouin zone and $s$-wave pairing is still energetically favorable. This results in a persisting magnitude of the $s$-wave order parameter in the surface states despite the energetic separation of surface bands.

\end{document}